\documentclass[useAMS,usenatbib,fleqn]{mn2e}
\usepackage{amstext}
\usepackage{amsfonts}
\usepackage{amssymb}
\usepackage{amsmath}

\usepackage{graphicx}
\usepackage{caption}
\usepackage{epstopdf}
\usepackage[varg]{txfonts}
\usepackage{xspace}
\usepackage{url}
\usepackage{multirow}
\usepackage{multicol} 
\usepackage{bm}		
\usepackage{pdflscape}	
\usepackage[deletedmarkup=sout]{changes}
\definechangesauthor[]{wek}
\usepackage{color}

\usepackage{times}

\newcommand{\goodne}{281}
\newcommand{\goodsw}{228}

\newcommand{\angstrom}{\mathring{A}}

\newcommand{\teff}{\ensuremath{T_\textrm{eff}}}
\newcommand{\re}{\ensuremath{R_\textrm{e}}}
\newcommand{\logg}{\ensuremath{\log(g)}}
\newcommand{\mh}{[M/H]}
\newcommand{\kms}{km\,s\ensuremath{^{-1}}}

\title[Stellar Populations between NSC and NSD]{Stellar Populations in the Transition Region of Nuclear Star Cluster and Nuclear Stellar Disc}
	\author[A.~Feldmeier-Krause]
	{A.~Feldmeier-Krause$^{1}$\thanks{E-mail: feldmeier@mpia.de} \\
$^{1}$Max Planck Institute for Astronomy, K{\"o}nigstuhl 17, 69117 Heidelberg, Germany }

\begin{document}

\date{Accepted 2022 April 25. Received 2022 April 08; in original form 2021 October 08}

\pagerange{\pageref{firstpage}--\pageref{lastpage}} \pubyear{year}

\maketitle

\label{firstpage}

\begin{abstract}
The Milky Way nuclear star cluster (NSC) is located within the nuclear stellar disc (NSD) in the Galactic centre. It is not fully understood if the formation and evolution of these two components are connected, and how they influence each other. We study the stellar populations in the transition region of NSC and NSD. We observed two $\sim$4.3\,pc$^2$ fields with the integral-field spectrograph KMOS (VLT), located at r$\sim$20\,pc (\textgreater 4\,\re) to the Galactic East and West of the NSC. 
We extract and analyse medium-resolution stellar spectra of \textgreater 200 stars per field. The data contain in total nine young star candidates. We use stellar photometry to estimate the stellar masses, effective temperatures, and spectral types of the young stars. The stars are consistent with an age of 4-6\,Myr, they may have formed inside the Quintuplet cluster, but were dispersed in dynamical interactions. 
Most stars in the two fields are red giant stars, and we measure their stellar metallicities [M/H] using full spectral fitting. We compare our [M/H] distributions to the NSC and NSD, using data from the literature, and find that the overall metallicity decreases from the central NSC, over the transition region, to the NSD. The steep decrease of [M/H] from the NSC to the region dominated by the NSD indicates that the two components have distinct stellar populations and formation histories. 
 
\end{abstract}

\begin{keywords}
Galaxy: centre; Stars: late-type, early-type; infrared: stars.
\end{keywords}

\section{Introduction}

The Galactic centre hosts three components that dominate the gravitational potential at different scales: The supermassive black hole Sgr A*, the nuclear star cluster (NSC), and the nuclear stellar disc (NSD). The supermassive black hole dominates the galactic potential in the central r$\sim$1--2\,pc \citep{2014A&A...570A...2F,2017MNRAS.466.4040F}, then the NSC dominates until $\sim$25--40\,pc, where the NSD takes over \citep{2002A&A...384..112L,2020MNRAS.499....7S}. It is an open question how these components are connected and influence each other.

NSCs are common in galaxies with similar properties as the Milky Way (MW). NSCs contain several tens of millions of stars, located in a small volume of the Galaxy. The MW NSC has a dynamical mass of about 2$\times10^7$\,M$_{\sun}$\space \citep{2014A&A...570A...2F,2017MNRAS.466.4040F}. It has an effective radius of \re= 4--5\,pc and  is flattened with an axis ratio $q$=0.7--0.8 \citep{2014A&A...566A..47S,2016ApJ...821...44F,2020A&A...634A..71G}.
The stars in the NSC belong to several stellar populations, with a dominant old population \citep[\textgreater 5--10\,Gyr,][]{2003ApJ...597..323B,2011ApJ...741..108P,2020A&A...641A.102S}, which contributes 75--80\% of the stellar mass. \cite{2020A&A...641A.102S} detect signatures of intermediate-age stars, formed about 3\,Gyr ago and contributing $\sim$15\% of the stellar mass, and stars formed $\sim$100\,Myr ago and contributing a few percent. Further, there is a cluster of young stars \citep[$\sim$6\,Myr,][]{2006ApJ...643.1011P} concentrated in the central 1\,pc of the NSC \citep{2015A&A...584A...2F}. 

The formation and growth of NSCs is a long-standing question that has been studied extensively \citep[for a review see][and references therein]{2020A&ARv..28....4N}. The two main scenarios that have been suggested are star cluster infall \citep{1975ApJ...196..407T} and in situ star formation \citep{1982A&A...105..342L}.
In the former case, globular or young massive star clusters form in the galaxy. Due to dynamical friction the clusters migrate to the centre, where they build up the NSC. This scenario was tested with semi-analytic models and $N$-body simulations \citep[e.g.][]{1993ApJ...415..616C,2000ApJ...543..620O,2011MNRAS.418.2697H,2011ApJ...729...35A,2012ApJ...750..111A,2013ApJ...763...62A,2015ApJ...806..220A,2016MNRAS.461.3620G,2020ApJ...901L..29A} 
and is expected to have contributed to the formation of NSCs. 
On the contrary, NSCs can form directly in the galaxy centre via in situ star formation \citep[e.g.][]{2004ApJ...605L..13M,2006ApJ...642L.133B,2007PASA...24...77B,2007A&A...462L..27S}. 
This requires that gas is brought to the galaxy centre, which may happen e.g. via bar-driven gas infall \citep{1990Natur.345..679S,2015MNRAS.446.2468E}, gas-rich mergers \citep{1994ApJ...437L..47M,2010MNRAS.407.1529H}, or magneto-rotational instabilities in gas discs \citep{2004ApJ...605L..13M}.

The currently favoured scenario is a combination of the two processes: Globular cluster infall appears to be the dominant process to form NSCs for galaxies with masses $\lesssim 10^9$ \,M$_{\sun}$, while in situ star formation dominates in more massive galaxies, such as the MW. Support for this scenario comes from observations of occupation fractions, metallicities, and star formation histories of NSCs and their host galaxies \citep[e.g.][]{2012ApJS..203....5T,2019ApJ...878...18S,2020A&ARv..28....4N,2021A&A...650A.137F}.

The MW NSC is surrounded by the NSD, a flattened stellar structure with a radius of $\sim$230\,pc \citep{2002A&A...384..112L}, a scale-height of $\sim$28--45\,pc \citep{2002A&A...384..112L,2013ApJ...769L..28N, 2022MNRAS.512.1857S}, and a mass of $\sim$7--10$\times10^8$\,M$_{\sun}$\space \citep{2020MNRAS.499....7S, 2022MNRAS.512.1857S}. It is unclear if the NSD is indeed a disc or rather a ring. 
One of the first studies of the NSD's stellar populations favoured a continuous star formation rate \citep{2004ApJ...601..319F}. Recently, \cite{2020NatAs...4..377N} analysed a larger data set, which supports several bursts of star formation. They found that \textgreater90\% of the stars are $\gtrsim$8\,Gyr old, and about 5\% formed $\sim$1\,Gyr ago. The past 500\,Myr had a continuous and low star formation rate, but there was a slight increase of the star formation rate $\sim$30\,Myr ago.

The formation of NSDs in disc galaxies appears to be driven by bars. They funnel gas to the central region of the galaxy, and new stellar structures form there. These structures can be NSDs, but also nuclear rings, inner bars or nuclear spiral arms \citep[e.g.][]{1983A&A...127..349A,1985A&A...150..327C, 1998MNRAS.298..267V,1991MNRAS.252..210B,2014MNRAS.445.3352C,2015MNRAS.446.2468E,2018MNRAS.481....2S,2019MNRAS.482..506G,2020A&A...643A..14G,2020A&A...643A..65B}. Galaxy mergers have also been suggested to trigger NSD formation \citep{2008MmSAI..79.1284M,2013MNRAS.429.3114C}, but the simulations cannot reproduce typical NSD sizes and kinematics \citep{2019MNRAS.482..506G,2020A&A...643A..14G}. 

As the NSC lies within the NSD, one can expect that there is some connection between these structures. The gas that feeds the NSC, and the star clusters that contributed to its growth, must have passed through the NSD. 
One way to learn more about the NSC-NSD connection, and how they influence each other, is to study and compare stellar populations and metallicity distributions in the NSC and NSD.

 Interstellar extinction in the Galactic centre is high, and for this reason metallicity measurements are usually made on $K$-band or $H$-band spectra of red supergiants or red giants. The latter are especially useful to derive the metallicity distribution of Galactic centre stars, as they are abundant, and sufficiently bright. Compared to supergiants, red giants are old enough to trace the dominant stellar populations of the NSC and NSD. 
Recently, \cite{2021A&A...649A..83F} sampled \textgreater3000 stars with the mid-resolution spectrograph KMOS in the NSD, and derived metallicities for $\sim$2700 stars. The sample includes stars with sub- to super-solar metallicities (-0.5 to +0.5\,dex). Most stars have super-solar metallicities, with a decreasing metallicity from the central disc to increasing scale height \citep{2021A&A...650A.191S}. Also stars in the NSC cover a wide range of metallicities from sub- to super-solar values. This was found both by high-spectral resolution studies of $\lesssim$15 stars \citep{2000ApJ...537..205R,2007ApJ...669.1011C,2017AJ....154..239R,2020ApJ...894...26T}, and medium-spectral resolution studies of \textgreater1000 stars \citep{2015ApJ...809..143D,2017MNRAS.464..194F,2020MNRAS.494..396F}, with super-solar mean metallicities.
In the transition region of NSC and NSD, 10--30\,pc away from the NSC centre, metallicities tend to be super-solar \citep{2007ApJ...669.1011C,2015A&A...573A..14R}, though sample sizes are too small ($\sim$10 stars) to measure the metallicity gradient from the NSC to the NSD. 

In this study, we analyse KMOS mid-resolution spectra of $\sim$500 stars, located in two fields $\sim$20\,pc distant from Sgr A*, on both sides of the Galactic plane. We study the stellar populations and detect nine young star candidates, for which we estimate the spectral types and stellar masses. Further, we measure the metallicities of red giant stars, and compare the metallicity distributions to those of the NSC \citep{2017MNRAS.464..194F,2020MNRAS.494..396F} and the NSD \citep{2021A&A...649A..83F}.

This paper is organized as follows: We describe the data set in Section \ref{sec:sec2}, and analyse the spectra in Section \ref{sec:sec3}. In Section \ref{sec:sec4} we present our results and compare them to other regions of the Galactic centre. Our discussion is presented in Section \ref{sec:sec5}, and we conclude in Section \ref{sec:sec6}.

\begin{figure*}
 \centering 
 \includegraphics[width=\textwidth]{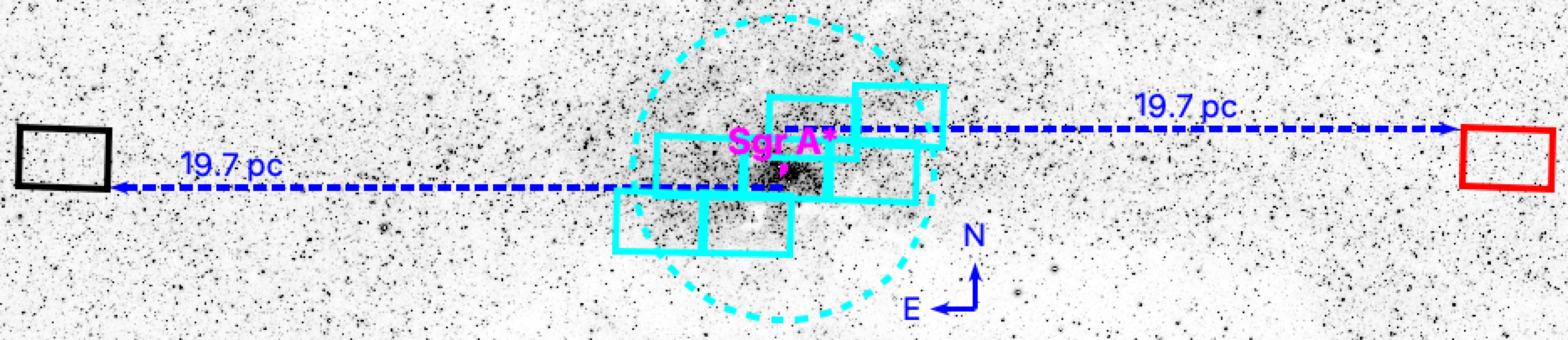}
 \caption{Locations of the KMOS mosaics, Galactic North is up, Galactic East is to the left. We observed two fields along the Galactic plane, at $\sim$20\,pc to the East (black) and West (red) of Sgr A* (magenta). 
 Seven further mosaic fields \citep[cyan,][]{2015A&A...584A...2F,2017MNRAS.464..194F,2020MNRAS.494..396F} cover the central nuclear star cluster; \re=4.2\,pc is shown as cyan dashed circle.
 The underlying $K_S$ band image is from the VVV Survey \citep{2012A&A...537A.107S}. } 
 \label{fig:observations}
\end{figure*}

\section[]{Data set}
\label{sec:sec2}
\subsection{Observations}
\label{sec:obs}
Our data were observed with KMOS on VLT-UT1 (Antu) 2014 June 7, and 2014 July 31 in service mode. KMOS \citep{2013Msngr.151...21S} is a multi-object spectrograph, it consists of 24 integral field units (IFUs). We used mosaic mode to arrange the IFUs in a close configuration, and fill the mosaic with 16 dithers. The mosaic covers 64.9\,arcsec $\times$ 43.3\,arcsec. The spatial sampling of KMOS is 0.2\,arcsec\,pixel$^{-1}$\,$\times$\,0.2\,arcsec\,pixel$^{-1}$.

We observed one mosaic field in the Galactic East, and one in the Galactic West of Sgr A*. Both fields are at a distance of about 20\,pc from Sgr A* and along the Galactic plane. The volume densities of NSC and NSD at this radius have the same order of magnitude \citep{ 2022MNRAS.512.1857S}. Figure~\ref{fig:observations} shows the regions that our data cover.
During the observations, three of the IFUs were inactive due to technical problems, which caused gaps in the mosaic fields. However, we observed each field twice, with different rotator angles, such that the gaps caused by inactive IFUs fall on different regions in each mosaic. Thus, we cover each field without gaps for at least one exposure, and for $\sim$75\% of each field we have even two exposures. Each exposure is analysed separately, and we measure stellar parameters on each exposure. We combine the stellar parameter measurements by taking the mean for each star.

Our spectra cover the 
 near-infrared $K$-band from $\sim$19\,340 to 24\,600\,$\angstrom$. They have a spectral resolution $R\sim$4000, and are sampled with $\sim$2.8\,$\angstrom$\,pixel$^{-1}$ along the dispersion axis. 
We observed a dark cloud (G359.94+0.17, located at $\alpha$\,$\approx$\,266\fdg2, $\delta$\,$\approx$\,$-$28\fdg9) for sky subtraction, and B-type dwarf stars for telluric correction. 
Table \ref{tab:observations} lists further details of our observations. 

\setcounter{table}{0}
\begin{table*}
 \centering
 \begin{minipage}{155mm}
\caption{Observation summary}
 \label{tab:observations}
\begin{tabular}{@{}cccccccccc@{}}
\noalign{\smallskip}
\hline
\noalign{\smallskip}
Field name& Mosaic number &Date&Exposure time &Seeing & Rotator Angle &Inactive KMOS IFU&R.A.&Dec.\\
&	&		& s &arcsec&degree&&degree&degree\\
 \noalign{\smallskip}
\hline
\noalign{\smallskip}
W	& 1& 2014 June 7 & 300& 0.6--1.0 &--60& 4,11,15 &266.328&	-29.130\\
	& 2&	 2014 July 31 & 300& 0.7--1.8 &120 & 4,11,15 &266.326&	-29.133\\
E	& 1& 2014 June 7 & 300& 0.7--1.1 &--60& 4,11,15 & 266.502	&-28.880\\	
	& 2&	 2014 July 31 & 300& 0.6--1.2 &120 & 4,11,15 &	266.500&	-28.883&\\	
 \hline 
\end{tabular}
\end{minipage}
\end{table*}

\subsection{Data Reduction}
\label{sec:reduction}
For data reduction we ultilised the KMOS pipeline \citep{2013A&A...558A..56D}, which is provided by ESO with EsoRex (ESO Recipe Execution Tool). We applied the standard recipes for dark subtraction, flat fielding, and wavelength calibration. For illumination correction we used the flat field exposures. We used B-type dwarf standard star spectra for telluric correction. Before the telluric correction, we removed stellar absorption lines and the blackbody spectrum of the standard star spectra with our own \textsc{idl} routine. 
Then, we reconstructed object and sky exposures separately to data cubes, and removed cosmic rays \citep{2001PASP..113.1420V,2013A&A...558A..56D}. For each observing block we have two sky exposures, which we combined to a mastersky frame. We scaled the sky to the object cubes before subtracting it, following \cite{2007MNRAS.375.1099D}. 

The spectral resolution of the KMOS data varies spatially, and for different IFUs \citep{2015ApJ...805..182G}. For this reason, we measured the line-spread function of the sky emission lines in \cite{2020MNRAS.494..396F}, and created a map of the spectral resolution for each KMOS IFU. We use those maps in our analysis.

For spectral extraction we used the $J$, $H$, $K_S$ photometric catalogue of the deep GALACTICNUCLEUS Survey \citep[GNS,][]{2019A&A...631A..20N}. This catalogue is several magnitudes deeper than our observations, and thus fairly complete for our magnitude ranges. We extracted only stars with $K_S\lesssim$16\,mag, as fainter stars are usually outshone by a nearby brighter star. The GNS lists $\sim$1700 such stars in the Eastern and $\sim$1400 stars in the Western field. In the few cases of an isolated $K_S$=16\,mag star, the S/N is rather low. 
For each star we used a circular aperture and summed the flux within the aperture. If the seeing was \textless1\arcsec, we used a 3 pixel radius (0\farcs6), else a 2 pixel radius (0\farcs4). The larger radius for good seeing helps to minimize the wavy pattern in the spectral continuum\footnote{{https://www.eso.org/sci/data-processing/faq/single-spaxel-spectra-show-wavelength-dependent-fluctuations-in-the-continuum.html}}. We subtracted the background flux using an annulus at 3-6\,pixel, or 2--5\,pixel if the star has a nearby neighbour star. We made sure that the extracted stars have no brighter neighbour stars within a distance of 0\farcs6, as these lead to inaccurate measurements. We extracted spectra of 650 (East) and 500 (West) stars, but later quality cuts (see Section \ref{sec:fitting}) decrease the number of stars we use for our analysis by more than half.

\section{Analysis}
\label{sec:sec3}
In this section, we describe the extinction correction and deselection of foreground stars. We measured spectral indices and describe our stellar parameter fitting.

\subsection{Extinction Correction and Foreground Star Contamination}
\label{sec:ext}
\begin{figure}
 \includegraphics[width=0.98\columnwidth]{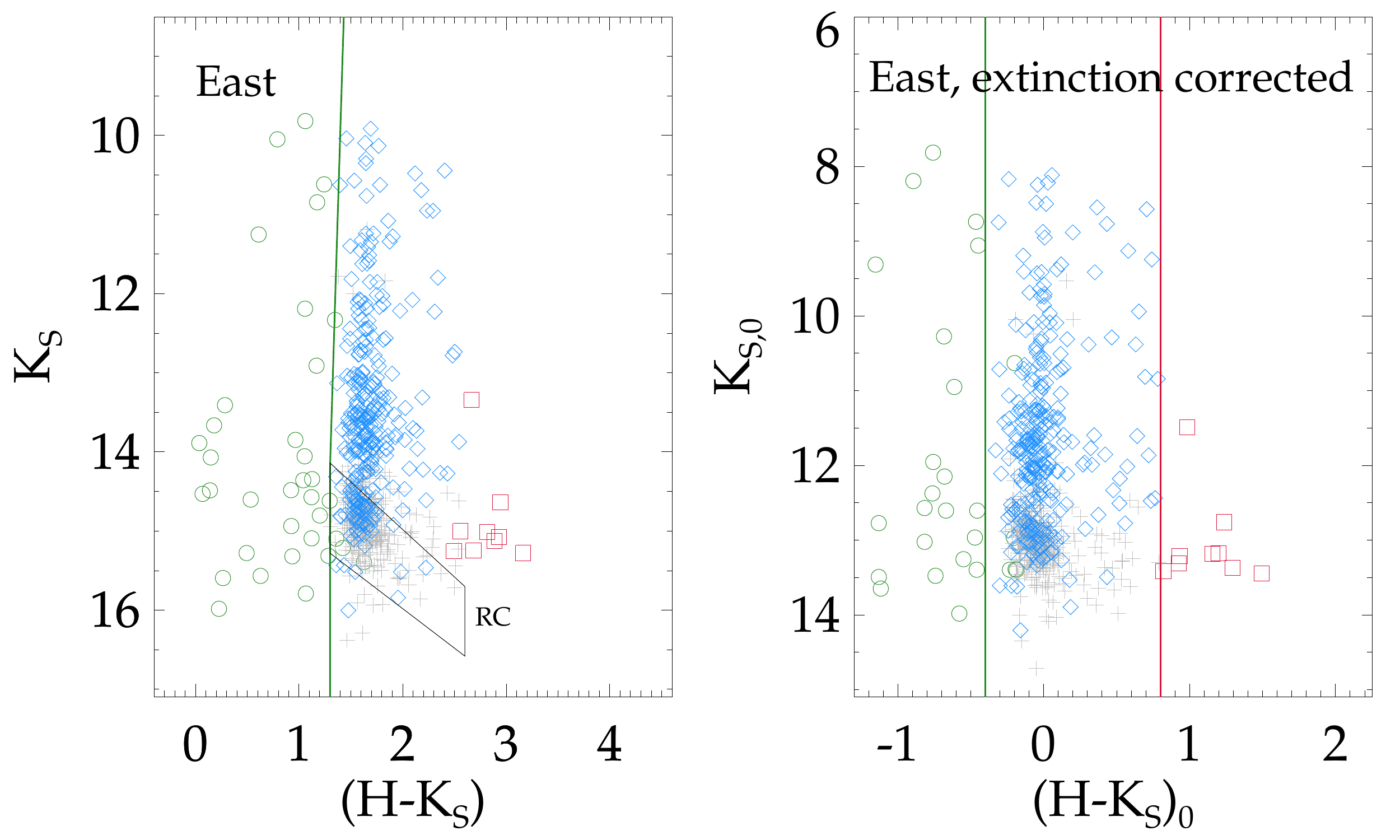}
 
 \includegraphics[width=0.98\columnwidth]{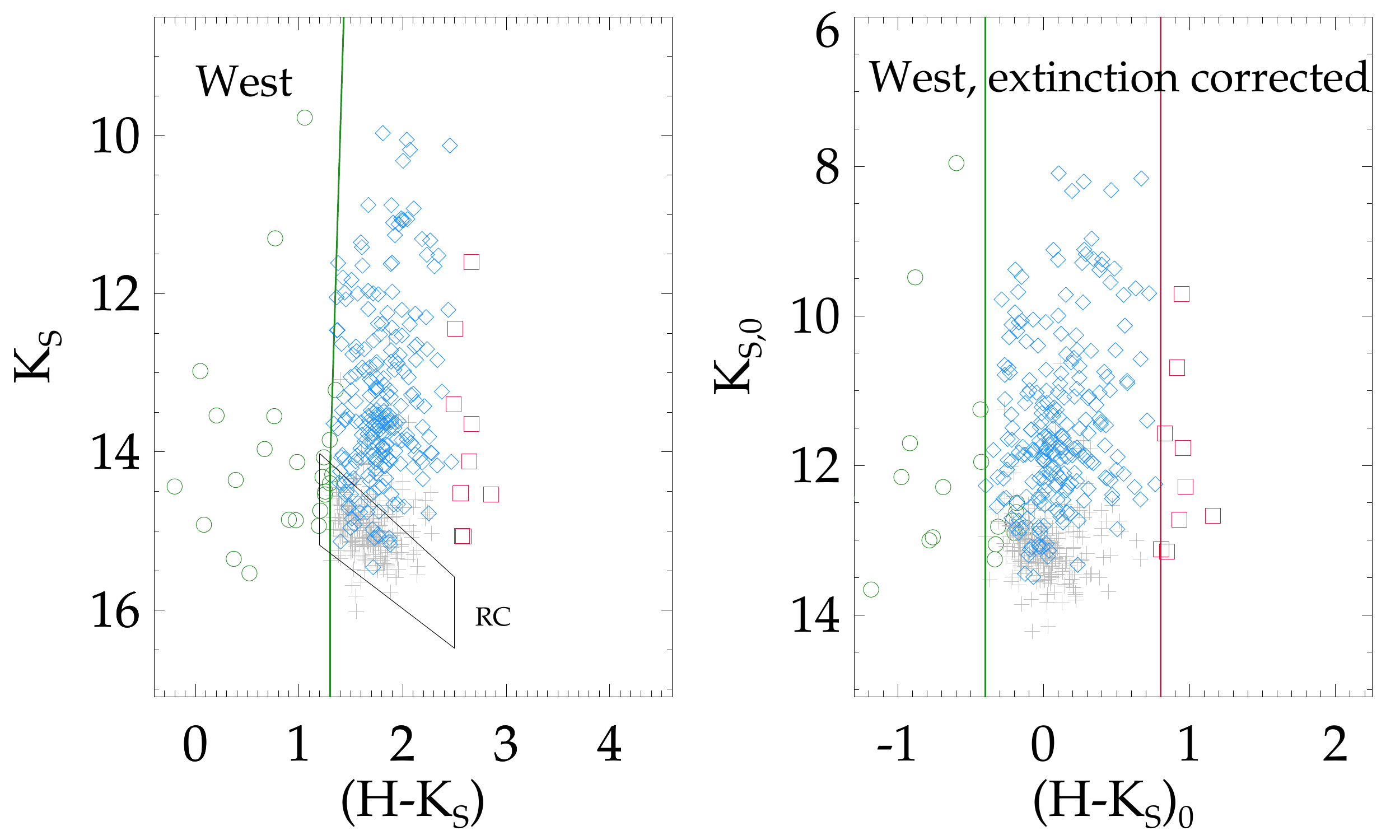}

 \caption{Colour-magnitude diagrams of stars in our two KMOS Fields, in the Galactic East (top, 650 stars), and Galactic West (bottom, 500 stars). We only show stars for which we extracted spectra and have $H$ and $K_S$ photometry. Left panels are before, right panels after extinction correction. Stars classified as member stars are denoted as blue diamond symbols, foreground stars as green circles, and background stars as red squares. The colour cuts used for classification are shown as red and green lines. Grey plus signs denote stars for which we extracted spectra, but later excluded them either because of nearby brighter stars or quality cuts (see Section \ref{sec:fitting}). The black parallelogram denotes the region we used to select red clump (RC) stars and construct the extinction maps (Section \ref{sec:ext}).}
 \label{fig:cmd}
\end{figure}

We used the GNS catalogue to create extinction maps, and followed the procedure outlined by \cite{2018A&A...610A..83N}. Briefly, 
we selected stars that are in the colour range ($H-K_S$) = 1.2 to 2.5\,mag for the Western field and ($H-K_S$) = 1.3 to 2.6\,mag for the Eastern field. The choice of the colour range was made after visual inspection of the colour magnitude diagrams in the two regions. Further, we made colour-dependent magnitude cuts to make sure we only consider red clump (RC) stars. The selected regions are shown as black parallelogram in Fig.~\ref{fig:cmd}. In each field, we have photometry for \textgreater2100 RC stars. We used Equation 5 of \cite{2018A&A...610A..83N} to derive the extinction $A_{K_S}$ for each individual RC star, assuming that the intrinsic colour of RC stars is ($H-K_S$)$_0$ =0.089\,mag, the filter effective wavelengths are $\lambda_H$=1.6506\,\micron, $\lambda_{K_S}$=1.21629\,\micron\space and the extinction coefficient $\alpha_{JHK_S}$=2.3. 
We created a map with the spatial sampling of our observations, 0\farcs2. For each pixel we used only stars within a radius of 12\arcsec, and computed a distance weighted mean extinction \citep[as in][]{2018A&A...610A..83N}. 

We applied the extinction map on the photometry of the stars for which we extracted spectra. Figure~\ref{fig:cmd} illustrates the colour-magnitude diagrams before (left) and after (right) extinction correction for the Field in the East (top) and West (bottom). Interestingly, the Eastern field has a more narrow range of colours than the Western field. This may indicate that extinction in the Eastern field is more uniform, with less spatial variation. 

To identify foreground stars we made colour cuts. This means we assume that the intrinsic $H-K_S$ colours of stars lie within a rather narrow range. This is the case for the NSC $H-K_S$=[$-$0.13\,mag, +0.38\,mag] \citep{2013ApJ...764..154D,2014CQGra..31x4007S}. As \cite{2021A&A...650A.191S} did for the NSD, we assumed that stars with colours bluer than ($H-K_S$) = max(1.3, $-0.0233\times K_S$+1.63)\,mag are foreground stars. In addition, we classified stars as foreground stars if their extinction corrected colours are bluer than ($H-K_S$)$_0$ = --0.4\,mag, and stars as background stars if they are redder than ($H-K_S$)$_0$ = 0.8\,mag, similar to \cite{2020MNRAS.494..396F}.

 \subsection{Spectral Indices}
\label{sec:index}
Spectral indices are useful to identify late-type stars and early-type stars. 
In order to measure spectral indices and identify young star candidates, we first derived the line-of-sight velocities by fitting the spectra with the \textsc{idl} program \textsc{pPXFv 5.2.1} \citep{2017MNRAS.466..798C} in the wavelength range 20\,880 to 23\,650\,$\angstrom$. As template spectra we used both late-type and early-type stars. As late-type stars we used the high resolution \cite{1996ApJS..107..312W} spectra, which we convolved to match the lower spectral resolution of KMOS. As early-type templates we used B-type dwarf spectra observed with KMOS as telluric standard stars. 
To estimate the uncertainties, we added random noise to the spectra and repeated the fits in 1000 realisations. The standard deviation of these 1000 fits is used as line-of-sight velocity uncertainty. We applied the line-of-sight velocity shift on the spectra and then measured spectral indices. 

 Early-type stars have weak or absent CO absorption, but tend to have Brackett (Br)~$\gamma$ absorption or even emission. In late-type stars, the CO equivalent width (EW) increases with decreasing effective temperature \teff. Also, CO lines are stronger for supergiant than for giant stars. The Na \sevensize{I} \normalsize
doublet is sensitive to the stellar metallicity \citep{2021A&A...649A..83F}. We measured the following spectral indices: First CO band head ($\sim$22\,935~$\angstrom$), Na \sevensize{I} \normalsize doublet (22\,062~$\angstrom$ and 22\,090~$\angstrom$), and Ca \sevensize{I} \normalsize triplet (22\,614~$\angstrom$, 22\,631~$\angstrom$, 22\,657~$\angstrom$)
(as defined by \citealt{2001AJ....122.1896F}); Br~$\gamma$ and H$_2$O indices (as defined by \citealt{2021A&A...649A..83F}). Asymptotic giant branch (AGB) stars have broad H$_2$O features (index \textless 0.7), and are rather bright ($K_{S,0}$\textless 7.5\,mag). We have two stars, one in each field, with H$_2$O index\textless 0.7, but with $K_{S,0}$\textgreater 8.1\,mag. Thus, the fraction of AGB stars in our sample is negligible.

\subsection{Stellar Parameter Fitting}
\label{sec:fitting}

We measured the stellar parameters total metallicity \mh, effective temperature \teff, and the surface gravity \logg\space on our KMOS spectra. 
We used full spectral fitting with the code 
 \textsc{StarKit} 
 \citep{2015zndo.....28016K}, which applies Bayesian sampling \citep[Multinest v3.10, pymultinest v2.11] {2008MNRAS.384..449F,2009MNRAS.398.1601F,2014A&A...564A.125B,2019OJAp....2E..10F}, and interpolates a grid of synthetic model spectra to find the best-fitting stellar parameters to an observed stellar spectrum.
The same code was used to constrain stellar parameters of NSC stars by \cite{2015ApJ...809..143D}, \cite{2017MNRAS.464..194F}, and \cite{2020MNRAS.494..396F}.

As in our previous work \citep{2017MNRAS.464..194F,2020MNRAS.494..396F}, we used the synthetic spectra of the PHOENIX spectral library \citep{2013A&A...553A...6H}. These spectra cover a grid with the following ranges and step sizes: \mh\space from $-$1.5\,dex to +1.0\,dex, $\Delta$\mh\space = 0.5\,dex, $\teff$\space from 2\,300 K to 12\,000 K, $\Delta\teff$\space = \,100\,K, $\logg$\space from 0.0\,dex to 6.0\,dex, $\Delta\logg$\space = 0.5\,dex. In addition to these parameters, we fit the line-of-sight velocity $v_z$.
For the latter we used the line-of-sight velocity measurement and uncertainty of Section \ref{sec:index} as Gaussian prior. 
For \mh\space and \teff\space we used uniform priors in the ranges mentioned above, for \logg\space we used information from the star's extinction corrected $K_{S,0}$ magnitude and CO equivalent width to restrict the uniform prior bounds as follows: Bright stars with $K_{S,0}$ $\leq$ 10\,mag and $EW_\text{CO}\geq$25\,$\angstrom$ have the bounds 0.0\,dex\,\textless\,$\logg$\,\textless 2.0\,dex; stars with $K_{S,0}$\textless 12\,mag or $EW_\text{CO}\geq$25\,$\angstrom$ have 0.0\,dex\,\textless\,$\logg$\,\textless 4.0\,dex, and all other stars have 2.0\,dex\,\textless\,$\logg$\,\textless 4.5\,dex. Potential foregrounds star (with blue $(H-K_S)_0$\textless$-0.5$\,mag) have 0.0\,dex\,\textless\,$\logg$\,\textless 6.0\,dex, regardless of $K_{S,0}$ and $EW_\text{CO}$.

For the PHOENIX model spectra, \mh\space stands for the overall metallicity of all elements, thus [$\alpha$/H] = \mh, and [$\alpha$/Fe] = 0.
Because of the low spectral resolution, we refrain from adding [$\alpha$/Fe] as fitting parameter. However, high spectral resolution studies have shown that stars in the NSC have a range of elemental abundances \citep{2020ApJ...894...26T}. For this reason we derived our systematic uncertainties using stars with a wide range of elemental abundances, see Appendix \ref{sec:xsl} for details.

Before the fit, the model spectra are convolved to match the spectral resolution of our data. For each KMOS spectrum, we used the spectral resolution at the respective location on the IFU, as it varies spatially (see Sect. \ref{sec:reduction}). We modelled the spectral continuum shape using a fifth degree polynomial function, and fit the spectra in the wavelength region $\lambda$\,=\,20\,900--\,22\,900\,$\angstrom$, excluding two regions ($\lambda$\,=\,22\,027--\,22\,125\,$\angstrom$ and $\lambda$\,=\,22\,575--\,22\,685\,$\angstrom$) with Na and Ca lines. This was also done by \cite{2017MNRAS.464..194F,2020MNRAS.494..396F} and is motivated by the strong Na \sevensize{I} \normalsize doublet and Ca \sevensize{I} \normalsize triplet features of Galactic centre stars compared to Galactic disc stars \citep{1996AJ....112.1988B,2017MNRAS.464..194F}. Excluding these regions prevents that \mh\space is biased to unrealistic high values, and means that our measurements are consistent to \cite{2017MNRAS.464..194F,2020MNRAS.494..396F}.

We made several quality cuts to ensure that the stellar parameters are obtained from spectra with sufficiently high quality and give precise measurements. In detail,
the selected spectra have a ratio of signal to residuum noise (S/rN)$\geq$ 10 of the \textsc{pPXF} fit (Section \ref{sec:index}), and a S/rN\textgreater 20 of the \textsc{starkit} fit. Further, we deselect stars with high fitting uncertainties $\sigma_{T_\text{eff}}\geq$250\,K, $\sigma_{\mh}\geq$0.25\,dex, $\sigma_{\logg}\geq$1\,dex, or $\sigma_{v_z}\geq$10\,\kms. 

If we had multiple spectra of the same star, which was the case for $\sim$45\% of the stars, we computed the average values of the individual stellar parameter measurements. If the standard deviation $\sigma_\text{sd}$ of these measurements exceeds the fitting uncertainties $\sigma_\text{fit}$, we used the standard deviation as statistical uncertainty. For each star with only one spectrum, we compared the fitting uncertainty $\sigma_\text{fit}$ to the median standard deviation of the stars with multiple exposures, and used which ever is greater as statistical uncertainty.

We estimated our systematic uncertainty by fitting 237 spectra of the X-SHOOTER spectral library (XSL) DR2 \citep{2014A&A...565A.117C,2020A&A...634A.133G} with \textsc{starkit}, see Appendix \ref{sec:xsl} for details. We compared our results with the stellar parameter measurements of \cite{2019A&A...627A.138A}, and use the standard deviation as systematic uncertainties. The exact values are $\sigma_{\Delta \mh}$=0.32\,dex, $\sigma_{\Delta \teff}$=272\,K, $\sigma_{\Delta \logg}$=1.0\,dex. 
These values are higher than the systematic uncertainties derived in \citet[][$\sigma_{\Delta \mh}$\,=\,0.24\,dex, $\sigma_{\Delta \teff}$\,=\,205\,K, and $\sigma_{\Delta \logg}$\,=\,1.0\,dex]{2017MNRAS.464..194F}.
We note that \cite{2017MNRAS.464..194F} used a smaller and inhomogeneous sample of reference stars, with a more narrow range of stellar parameters.

We added the statistical and systematic uncertainties in quadrature to obtain our final total stellar parameter uncertainties. The total uncertainties are dominated by the systematic uncertainties. 
The average total uncertainties are $\sigma_{\mh}$\,=\,0.36\,dex, $\sigma_{\teff}$\,=\,295\,K, and $\sigma_{\logg}$\,=\,1.05\,dex. Due to the rather large systematic uncertainties we refrain from analysing the \teff-\logg-diagram of our sample of stars. Instead, we compare \mh\space distributions in different regions of the NSC and NSD in Section \ref{sec:profile}.

\begin{figure}
 \includegraphics[width=0.98\columnwidth]{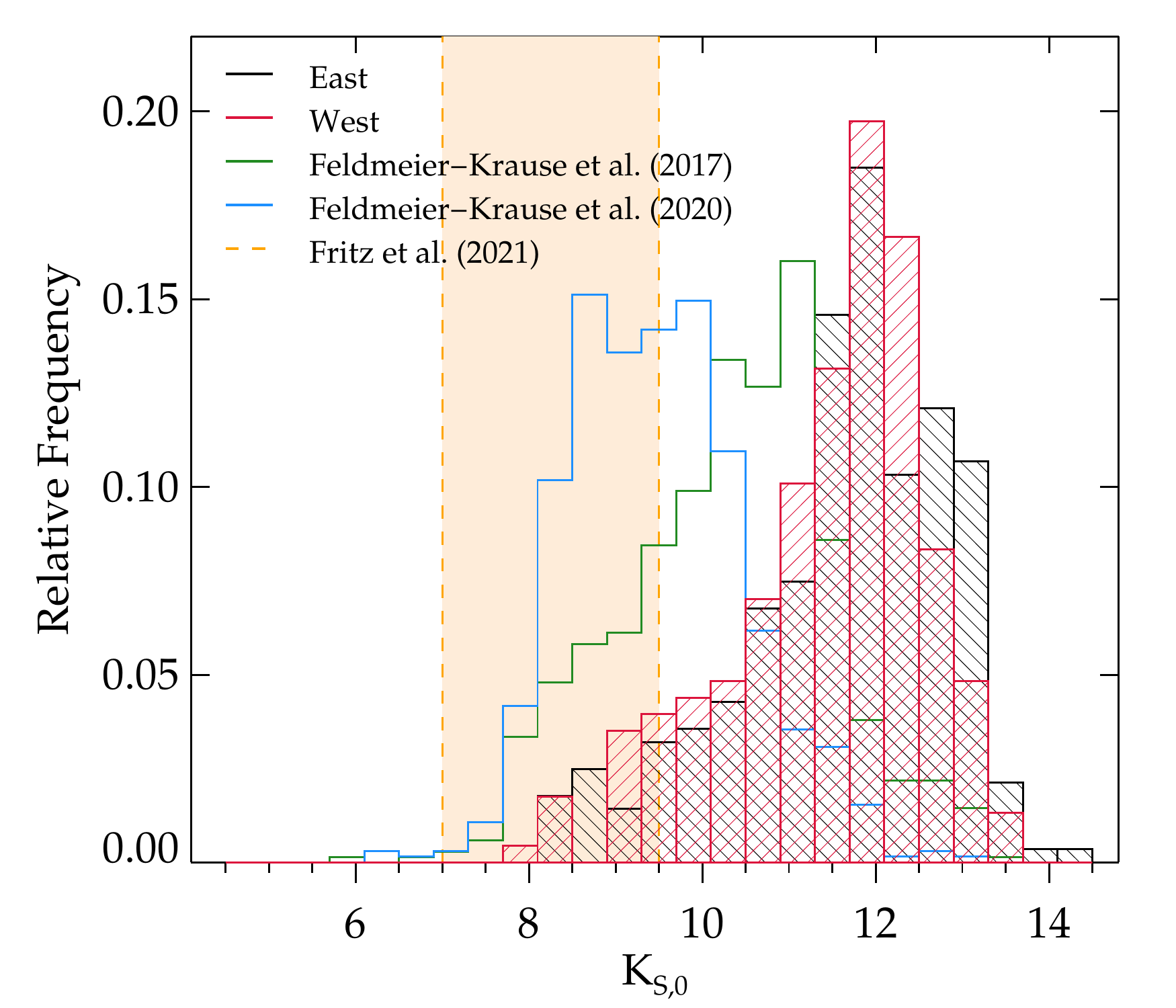}
 
 \caption{The histograms of $K_{S,0}$ (extinction corrected), for the East (black), and West (red) fields are similar. \citet[green]{2017MNRAS.464..194F}, \citet[blue]{2020MNRAS.494..396F}, and \citet[orange]{2021A&A...649A..83F} sample brighter stars. }
 \label{fig:ks0}
\end{figure}

\begin{table*}
 \centering
 \begin{minipage}{155mm}
\caption{Completeness of data set in apparent magnitude $K_S$, and in absolute magnitude $M_{K_S}$ (assuming a distance of 8.178\,kpc, which is the distance to Sgr~A*, \citealt{2019A&A...625L..10G}). We list the magnitudes at a completeness of 80\%, 50\%, and 20\%.}
 \label{tab:completeness}
\begin{tabular}{@{}cccccccc@{}}
\noalign{\smallskip}
\hline
\noalign{\smallskip}
Field name& 80\% completeness $K_S$& 50\% $K_S$ & 20\% $K_S$ &80\% $M_{K_S}$& 50\% $M_{K_S}$ & 20\% $M_{K_S}$\\
&	mag&mag&mag&	mag&mag&mag\\
\hline
E & 13.1$\pm$1.3 & 14.07$\pm$0.13 & 14.98$\pm$0.12 &-1.46& -0.49& 0.42\\
W & 13.20$\pm$0.13 & 14.08$\pm$0.11 & 14.97$\pm$0.12 &-1.36& -0.48& 0.41\\
 \hline 
\end{tabular}
\end{minipage}
\end{table*}

\subsection{Completeness}
\label{sec:comp}
Before we can compare our measurements with other studies in the literature, we need to know how deep our observations are in comparison to these studies. We do this by estimating the completeness, defined as the fraction of stars we observed at a given magnitude. 

As reference for the actual number of stars in the field at a given magnitude we used the photometric GNS catalogue, which is several magnitudes deeper than our observations, and can be considered as nearly 100\% complete in the magnitude range of interest. We compared the number of stars for which we measured stellar parameters (after our quality cuts, Section \ref{sec:fitting}) to the number of stars in the GNS catalogue, binned by magnitude. We used different $K_S$ magnitude bin widths (0.7, 1.0, and 1.2\,mag) and minimum $K_S$ values. Then, we computed at which $K_S$ magnitude our data are 80\%, 50\%, or 20\% complete. The results are listed in Table \ref{tab:completeness}, the uncertainties are the standard deviations for different magnitude binning. We note that the completeness values are averages over each field. There may be variations within each field, for example due to inactive IFUs or different levels of crowding. 

The completeness ranges of the two fields are in agreement within their uncertainties. This is expected, since the observations were taken during the same nights with the same exposure times. There were only small changes in the seeing during the observations (see Table \ref{tab:observations}). However, the 80\% completeness of the Eastern field has a large uncertainty. This is because also at brighter magnitude bins, the maximum completeness scatters around 75-90\%, and never reaches 100\%. In the Western field, the completeness increases further for brighter stars, reaching $\sim$95\%. Both fields sample stars along the red giant branch, down to the red clump ($K_{S,0}\sim$13\,mag, $K_S\sim$14--16\,mag).

Our data reach their 50\% completeness at $K_S$=14.1\,mag. This is deeper than the KMOS observations of \cite{2017MNRAS.464..194F}, which reach 50\% completeness at 13.7\,mag, and \cite{2020MNRAS.494..396F}, which reach it at 11.9--12.5\,mag, depending on the field. 
This is due to the shorter exposure times (100--190\,s instead of 300\,s) and more crowded fields, within 1\,\re\space of the NSC, which makes it hard to reach the red clump ($K_{S,0}\sim$13\,mag) without adaptive optics. We compare the extinction corrected $K_{S,0}$ distributions in Fig.~ \ref{fig:ks0}. 
Our data are also deeper than \cite{2021A&A...649A..83F}, who sampled stars in the NSD with extinction corrected magnitudes $K_{S,0}$=7--9.5\,mag. Assuming an average extinction of $A_{K_S}$=1.85\,mag in our fields, we have 80\% completeness at $K_{S,0}$=11.3\,mag. Thus, we have high completeness in the magnitude range of \cite{2021A&A...649A..83F}, but there are only few stars of that brightness in our fields. 


\section{Results}
\label{sec:sec4}


\subsection{Young Star Candidates}

Young stars trace recent star formation events in the Galactic centre. In the NSC, there is a cluster of young stars located in the central r=0.5\,pc, the NSD hosts two young star clusters, Arches and Quintuplet, and in addition several isolated young stars. Here, we identify young star candidates in our data and classify them. 

\begin{figure}
 \centering 
 \includegraphics[width=0.99\columnwidth]{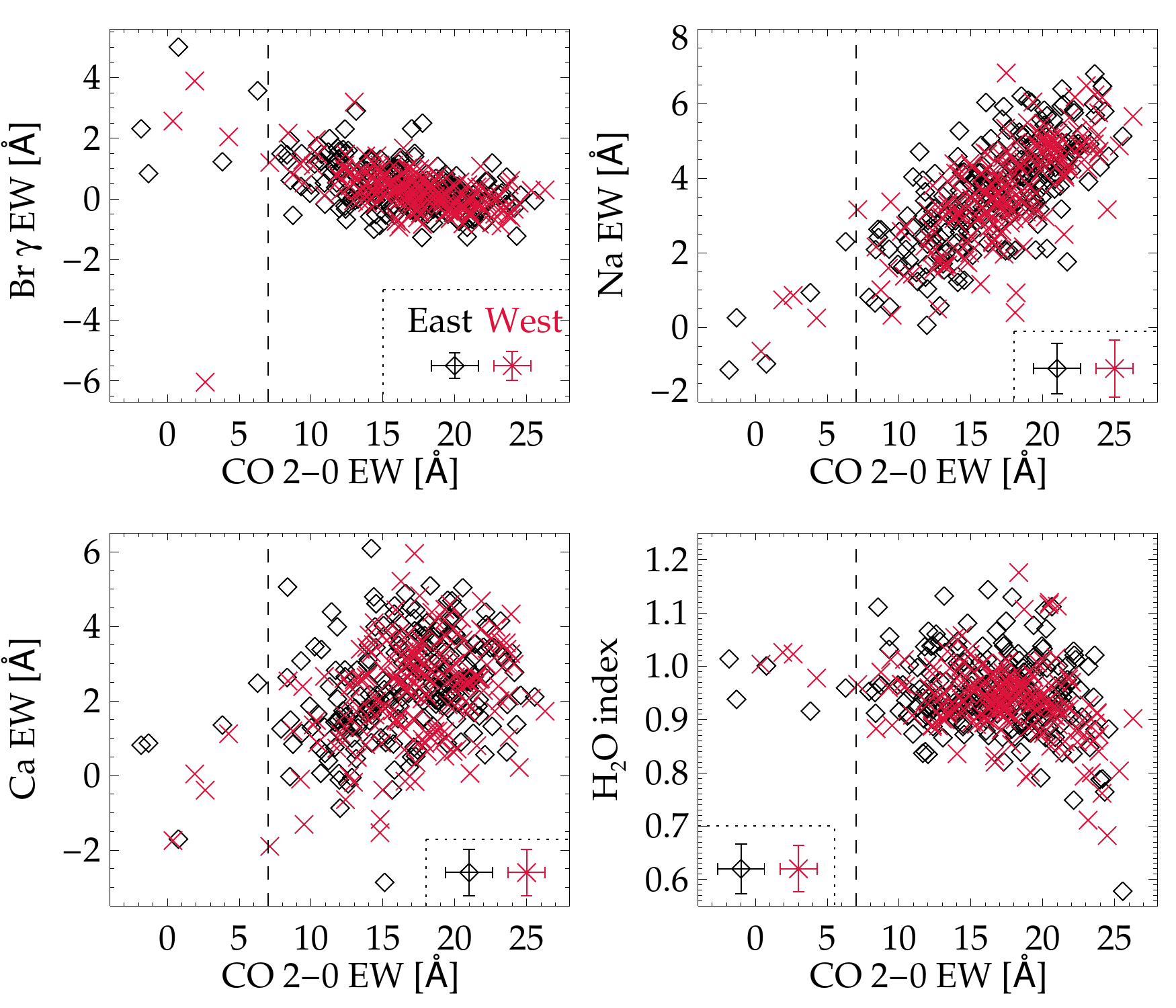}
 \caption{Spectral index measurements of our sample of stars, black for stars in the East field, red for West. We show $EW_{\text{Br}~\gamma}$, $EW_\text{Na}$, $EW_\text{Ca}$, and the H$_2$O index as function of $EW_\text{CO}$. The average uncertainties are shown in the bottom of the plots. The vertical line at $EW_\text{CO}$=7\,$\angstrom$ marks stars with low $EW_\text{CO}$; they are young star candidates.
 }
 
 \label{fig:indices}
\end{figure}

We use the spectral indices measured in Section \ref{sec:index} to identify hot young star candidates. The first CO band head is a proxy for the stellar effective temperature \citep[e.g.][]{1996AJ....112.1988B}, such that a lower value of $EW_\text{CO}$ indicates a higher \teff. Younger stars tend to be hotter and therefore have lower $EW_\text{CO}$ \citep{2018A&A...609A.109N}. 
Our index measurements are presented in Fig.~\ref{fig:indices}. As a function of $EW_\text{CO}$ we show, from left to right and top to bottom, $EW_{\text{Br} \gamma}$, $EW_\text{Na}$, $EW_\text{Ca}$, and the H$_2$O index. Stars with low $EW_\text{CO}$ also have rather low $EW_\text{Na}$ and $EW_\text{Ca}$.

\begin{figure}
 \centering 
 \includegraphics[width=0.99\columnwidth]{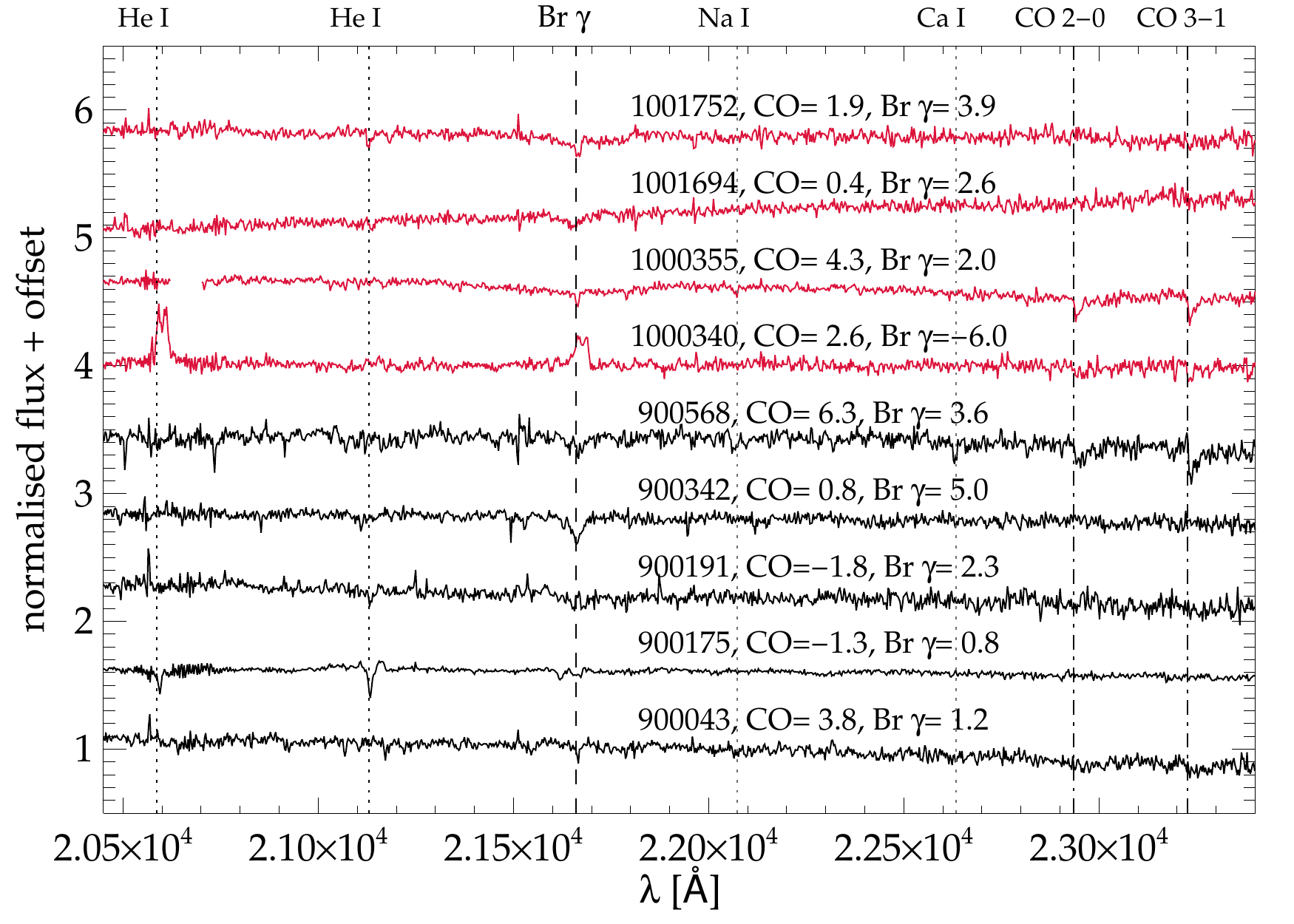}
 \caption{Spectra of stars with low CO absorption ($EW_\text{CO}$\textless7\,$\angstrom$). Black colour denotes stars in the East field, red in the West field. The spectra are shifted to rest-wavelength, normalised and an offset is added. 
 We show several spectral features with vertical lines, labelled on top of the plot. }
 \label{fig:lowcospec}
\end{figure}

We classify stars as young star candidates if their spectra have $EW_\text{CO}$\textless7\,$\angstrom$, we list them in Table \ref{tab:candindex}, and show them in Fig.~\ref{fig:lowcospec}. Young stars can also have Br~$\gamma$ either in absorption or in emission. We find that most of the stars with low $EW_\text{CO}$ have Br~$\gamma$ in absorption (\textgreater 0\,$\angstrom$), only one star (Id 1000340) has Br~$\gamma$ emission (\textless0\,$\angstrom$). Some stars also have He \sevensize{I} \normalsize features at 2.058\,\micron\space and 2.113\,\micron.

\begin{table}
 \centering
\caption{Candidate young stars with low CO absorption ($EW_\text{CO}$\textless7\,$\angstrom$)}
 \label{tab:candindex}
\begin{tabular}{ccccc}
\hline
ID& RA &Dec& $EW_\text{CO}$ & $EW_{\text{Br} \gamma}$ \\
&degree&degree&$\angstrom$&$\angstrom$\\
\hline
 900043& 266.50940& -28.87711& 3.8& 1.2\\
 900175& 266.49762& -28.88081& -1.3& 0.8\\
 900191& 266.49948& -28.88537& -1.8& 2.3\\
 900342& 266.50360& -28.88473& 0.8& 5.0\\
 900568& 266.49362& -28.88634& 6.3& 3.6\\ 
1000340& 266.33267& -29.12937& 2.6& -6.0\\
1000355& 266.33066& -29.12546& 4.3& 2.0\\
1001694& 266.32382& -29.13659& 0.4& 2.6\\
1001752& 266.32666& -29.13736& 1.9& 3.9\\
\hline 
 \end{tabular}
\end{table}

\begin{table*}
 \centering
 \begin{minipage}{155mm}
\caption{Photometry of the candidate young stars. $JHK_S$ photometry from \citet{2019A&A...631A..20N}; 3.6, 4.5, 5.8 and 8.0\,\micron\space photometry from \citet{2009yCat.2293....0S}, except for Id 900342, which has photometry from \citet{2008ApJS..175..147R}.}
 \label{tab:peculiars}

\begin{tabular}{@{}ccccccccl@{}}
\noalign{\smallskip}
\hline
\noalign{\smallskip}
ID& $J$ &$H$ & $K_S$ & 3.6\,\micron& 4.5\,\micron&5.8\,\micron&8.0\,\micron&Match \\
&mag&mag&mag&mag&mag&mag&mag&within 1\arcsec\\
 \noalign{\smallskip}
\hline
\noalign{\smallskip}
 900043& 18.04 $\pm$ 0.009& 15.14 $\pm$ 0.023& 13.72 $\pm$ 0.018&...&...&...&...&\\
 900175& 14.73 $\pm$ 0.011& 12.02 $\pm$ 0.007& 10.62 $\pm$ 0.010& 9.47 $\pm$ 0.05& 9.15 $\pm$ 0.04& 8.95 $\pm$ 0.06& 9.65 $\pm$ 0.14& \cite{2011MNRAS.417..114D}\\
 900191& 15.87 $\pm$ 0.015& 13.16 $\pm$ 0.012& 11.78 $\pm$ 0.007& 10.58 $\pm$ 0.05& 9.96 $\pm$ 0.06& 9.06 $\pm$ 0.04& 7.33 $\pm$ 0.08&\cite{2011ApJ...736..133A}\\
900342& 18.12 $\pm$ 0.011& 15.27 $\pm$ 0.014& 13.79 $\pm$ 0.009& 11.93 $\pm$ 0.04& 11.66 $\pm$ 0.05&...&...& \\
900568& 19.19 $\pm$ 0.021& 15.92 $\pm$ 0.009& 14.23 $\pm$ 0.009&...&...&...&...& \\
 
1000340& 17.72 $\pm$ 0.009& 15.08 $\pm$ 0.023& 13.70 $\pm$ 0.018&...&...&...&...&\cite{2011MNRAS.417..114D}\\
1000355& 17.84 $\pm$ 0.011& 14.98 $\pm$ 0.007& 13.64 $\pm$ 0.010&...&...&...&...&\\
1001694& ...& 15.69 $\pm$ 0.012& 13.63 $\pm$ 0.007& 11.33 $\pm$ 0.09& 10.09 $\pm$ 0.08& 8.62 $\pm$ 0.04& 7.42 $\pm$ 0.03& \cite{2021AA...651A..88N}\\
1001752& ...& 15.89 $\pm$ 0.014& 14.34 $\pm$ 0.009&...&...&...&...&\\
\hline 
\end{tabular}
\end{minipage}
\end{table*}


\begin{table*}
 \centering
 \begin{minipage}{155mm}
\caption{Properties of OB type candidates, derived from $HK_S$ photometry and assuming an intrinsic colour $H-K_S\sim$--0.09\,mag. Spectral type estimates are derived from the \teff\space range, except for Id 900175, for which we compared the spectrum to reference spectra. While this star is likely a supergiant, all other stars are in the magnitude range of dwarf stars.}
 \label{tab:obtab}
\begin{tabular}{@{}c|cc|ccllllll@{}}
 \hline
&\multicolumn{2}{c}{Extinction map}&\multicolumn{2}{c}{Instrinsic colour}\\
Id & $K_{S,0}$& $A_{K_S}$ & $K_{S,0}$& $A_{K_S}$&Mass &\teff&\logg&log($L$)&Spectral type&$V_\text{los}$\\
 & mag& mag & mag& mag&M$_{\sun}$ &K&dex&L$_{\sun}$&&\kms\\
\hline
 900043 & 11.80& 1.92 & 12.00$\pm$ 0.22 & 1.72$\pm$ 0.20 & 14$^{+ 2}_{- 2}$ & 30000$^{+ 3000}_{- 2500}$ & 3.99$^{+ 0.08}_{- 0.08}$ & 4.4$^{+ 0.2}_{- 0.2}$&O9--B1& -9.0$\pm$ 5.0\\
 900175 & 8.75& 1.87 & 8.88$\pm$ 0.21 & 1.74$\pm$ 0.20 & 60$^{+27}_{-33}$ & 26500$^{+ 14000}_{- 5500}$ & 2.95$^{+ 0.32}_{- 0.32}$ & 5.3$^{+ 0.5}_{- 0.2}$&O8--O9.5& 59.6$\pm$ 55.5\\
 900191 & 10.05& 1.73 & 10.07$\pm$ 0.21 & 1.71$\pm$ 0.20 & 38$^{+13}_{-15}$ & 37000$^{+ 6000}_{- 9000}$ & 3.45$^{+ 0.30}_{- 0.18}$ & 5.1$^{+ 0.4}_{- 0.2}$&O4--B0.5& 23.1$\pm$ 77.5\\
 900342 & 11.91& 1.88 & 12.00$\pm$ 0.22 & 1.79$\pm$ 0.21 & 14$^{+ 2}_{- 2}$ & 30000$^{+ 3000}_{- 2500}$ & 3.99$^{+ 0.08}_{- 0.08}$ & 4.4$^{+ 0.2}_{- 0.2}$&O9--B1& 145.0$\pm$ 37.2\\
 900568 & 12.32& 1.91 & 12.20$\pm$ 0.24 & 2.03$\pm$ 0.23 & 13$^{+ 2}_{- 2}$ & 29000$^{+ 3000}_{- 2500}$ & 4.03$^{+ 0.08}_{- 0.08}$ & 4.3$^{+ 0.2}_{- 0.2}$&O9.5--B1& -16.2$\pm$ 16.2\\
1000355 & 11.79& 1.85 & 12.01$\pm$ 0.20 & 1.63$\pm$ 0.19 & 14$^{+ 2}_{- 2}$ & 30000$^{+ 3000}_{- 2500}$ & 3.99$^{+ 0.08}_{- 0.06}$ & 4.4$^{+ 0.2}_{- 0.2}$&O9--B1& -96.9$\pm$ 5.4\\
1001694 & 11.67& 1.96 & 11.16$\pm$ 0.28 & 2.47$\pm$ 0.27 & 19$^{+ 9}_{- 4}$ & 34000$^{+ 5000}_{- 4000}$ & 3.85$^{+ 0.12}_{- 0.14}$ & 4.8$^{+ 0.2}_{- 0.2}$&O6--B0.5& -54.9$\pm$ 47.7\\
1001752 & 12.52& 1.82 & 12.47$\pm$ 0.22 & 1.87$\pm$ 0.21 & 12$^{+ 2}_{- 2}$ & 28000$^{+ 2000}_{- 3000}$ & 4.05$^{+ 0.08}_{- 0.06}$ & 4.1$^{+ 0.2}_{- 0.2}$&B0--B1.5& -78.1$\pm$ 67.0\\
\hline 
\end{tabular}
\end{minipage}
\end{table*}

We list the young star candidates in Table \ref{tab:peculiars}, with their $JHK_S$ \citep{2019A&A...631A..20N} and IRAC 3.6, 4.5, 5.8, and 8.0\,\micron\space \citep{2008ApJS..175..147R,2009yCat.2293....0S} photometry. Three of the stars have $H$, $K_S$, and 8.0\,\micron\space photometry, thus we can check if they fulfill the colour criterion suggested by \cite{2018A&A...609A.109N} to separate young stellar objects (YSOs) and OB-type stars from late-type stars. The younger objects have a higher $H-$[8.0] colour than late-type stars at the same value of $H-K_S$. This is indeed the case for ID 900191 and 1001694, with $H-$[8.0]\textgreater 5.8\,mag. But Id 900175 has $H-$[8.0] =2.4\,mag, similar to late-type stars, though it has no CO, Na, or Ca absorption but instead Br~$\gamma$ and He \sevensize{I} \normalsize in absorption (see Fig.~\ref{fig:lowcospec}). 
This star is also a Paschen $\alpha$ emission candidate \citep{2011MNRAS.417..114D}, as is Id 1000340, which has emission at Br~$\gamma$ and He \sevensize{I} \normalsize (2.058\,\micron), and may be a Wolf Rayet (WR) star. 
The fraction of young star candidates over the total number of stars that satisfy our quality cuts is similar for the East and West fields.

Since we lack mid-infrared data, we cannot differentiate whether the young star candidates are 2--8 Myr old OB-type stars, similar to those found in the central parsec of the NSC, and in the Arches and Quintuplet clusters \citep{1999ApJ...514..202F,2004ApJ...611L.105N,2006ApJ...643.1011P}, or rather \textless1\,Myr old YSOs. However, we compared the position of the stars in the $(H-K_S)_0$ vs. $K_{S,0}$ colour magnitude diagram with the mean values of dwarf stars \citep{2012ApJ...746..154P,2013ApJS..208....9P}\footnote{\url{https://www.pas.rochester.edu/~emamajek/EEM_dwarf_UBVIJHK_colors_Teff.txt}, Version 2021.03.02}, and found that they lie within the typical ranges of O to early B dwarf stars. Only Id 900175 is about 1\,mag brighter than an O3V star, and therefore a giant or supergiant star candidate. O/B-type stars have \teff\space$\sim$10\,000--36\,000\,K, which is higher than the temperature grid of the PHOENIX spectra used in Section \ref{sec:fitting}.

We used the $H$ and $K_S$ photometry to estimate the stellar mass, \teff, \logg, and luminosity log($L$), assuming that the stars (except the possible WR stars Id 1000340) are O/B type stars. We proceed in a similar way as \cite{2015A&A...584A...2F}. 
The $(H-K_S)_0$ colours of the OB star candidates derived from the extinction map have a large spread, from --0.35 to 0.27\,dex. This is because the extinction map was derived by averaging over several stars. Here, we used the intrinsic colour to correct each OB star candidate.
As intrinsic colour we took the mean dwarf stellar colours of \cite{2012ApJ...746..154P,2013ApJS..208....9P}, 
converted from 2MASS to ESO filters \citep{2001AJ....121.2851C}\footnote{\url{https://irsa.ipac.caltech.edu/data/2MASS/docs/releases/allsky/doc/sec6_4b.html}}, extrapolated for stars hotter than O9V, for which there are no $HK_S$ but only $UBV$ photometry, and interpolated to a fine grid. The intrinsic colours are about --0.09\,mag. 
We shifted the photometry of the observed OB stars to match the intrinsic colour of the mean star grid, assuming a power-law for the extinction $A_\lambda \propto \lambda^{-\alpha}$, with $\alpha$=2.23$\pm$0.07 \citep{2020A&A...641A.141N}. We list the extinction corrected magnitude $K_{S,0}$ and $A_{K_S}$ for the stars in Table \ref{tab:obtab}, when using the extinction map and when using the intrinsic colour.

Next, we used the intrinsic colour $(H-K_S)_0$ and shifted $K_{S,0}$ to compute the likelihood that a star has a certain mass, \teff, \logg, and log($L$). We used the PARSEC version 1.2S isochrones \citep{2012MNRAS.427..127B,2015MNRAS.452.1068C,2017ApJ...835...77M,2019A&A...632A.105C} with a \cite{2001MNRAS.322..231K} initial mass function and ages 1-8\,Myr (with steps of 0.25\,Myr), and \mh\space = --0.2 to 0.6\,dex (with steps of 0.1\,dex). From these isochrones, we selected stars with 8000\,K\textless\teff\textless 52,000\,K, which includes O/B type stars. We converted the isochrone $(H-K_S)_\text{iso}$ and $K_{S,\text{iso}}$ from 2MASS to ESO photometry, and for every star we computed the likelihood $\mathcal{L}$ that the observed $(H-K_S)_0$ equals $(H-K_S)_\text{iso}$ and $K_{S,0}$ equals $K_{S,\text{iso}}$ as 
\begin{eqnarray}
\mathcal{L}=\frac{1}{\sqrt{{2\pi}}\sigma_{K_{s,0}}}\exp\left ( {-\frac{1}{2} \left( \frac{K_{S,0} - K_{S,\text{iso}}}{\sigma_{K_{s,0}}}\right)^2} \right ) \times \nonumber
\\
\hspace{5mm}\frac{1}{\sqrt{{2\pi}}\sigma_{(H-Ks)_0}}\exp \left ({-\frac{1}{2} \left( \frac{(H-K_S)_{0} - (H-K_S)_\text{iso}}{\sigma_{(H-Ks)_0}}\right)^2} \right ), 
\label{eq:lik}
\end{eqnarray}
where we assumed $\sigma_{(H-Ks)_0}$=0.05\,dex \citep[following][]{2015A&A...584A...2F}. 
Each isochrone point is associated with a stellar property such as stellar mass, \teff, \logg, and log($L$). From the likelihoods we computed the probability functions for each stellar property. We report the median mass, \teff, \logg, and log($L$) for each OB star candidate in Table \ref{tab:obtab}, the uncertainties denote the 0.16 and 0.84 percentiles. 

From the range of \teff\space and the mean dwarf star \teff\space \citep{2012ApJ...746..154P,2013ApJS..208....9P}, we also estimated the range of possible spectral types. The stars are in the ranges of late O- to early B-type main sequence stars. Our spectral type and mass for Id 1001694 are in agreement with the radio continuum estimates by \citet[B0--B0.5, 14.4\,M$_{\sun}$] {2021AA...651A..88N}. 
We compared our stellar spectra to higher resolution spectra of OB type stars \citep{2005ApJS..161..154H,2018ApJS..238...29P,2020A&A...634A.133G}, convolved to KMOS resolution, to test our spectral classification. 
Our visual inspection shows good agreement: Most stars are in the range O9V-B1.5V. Undetectable He\,{\sc ii}\space \normalsize 2.189\,\micron\space absorption excludes earlier O types, whereas later B and A type stars would require stronger Br\,$\gamma$ absorption. The stars 900191 and 1001752 have indications of He\,{\sc i}\space \normalsize absorption at 2.113\,\micron, which is also found in late O and early B dwarf stars \citep{2005ApJS..161..154H,2018ApJS..238...29P}. However, a higher S/N is required to clearly detect the features at 2.113\,\micron\space and 2.189\,\micron, and obtain a precise classification of these stars. 

\begin{figure}
 \centering 
 \includegraphics[width=0.99\columnwidth]{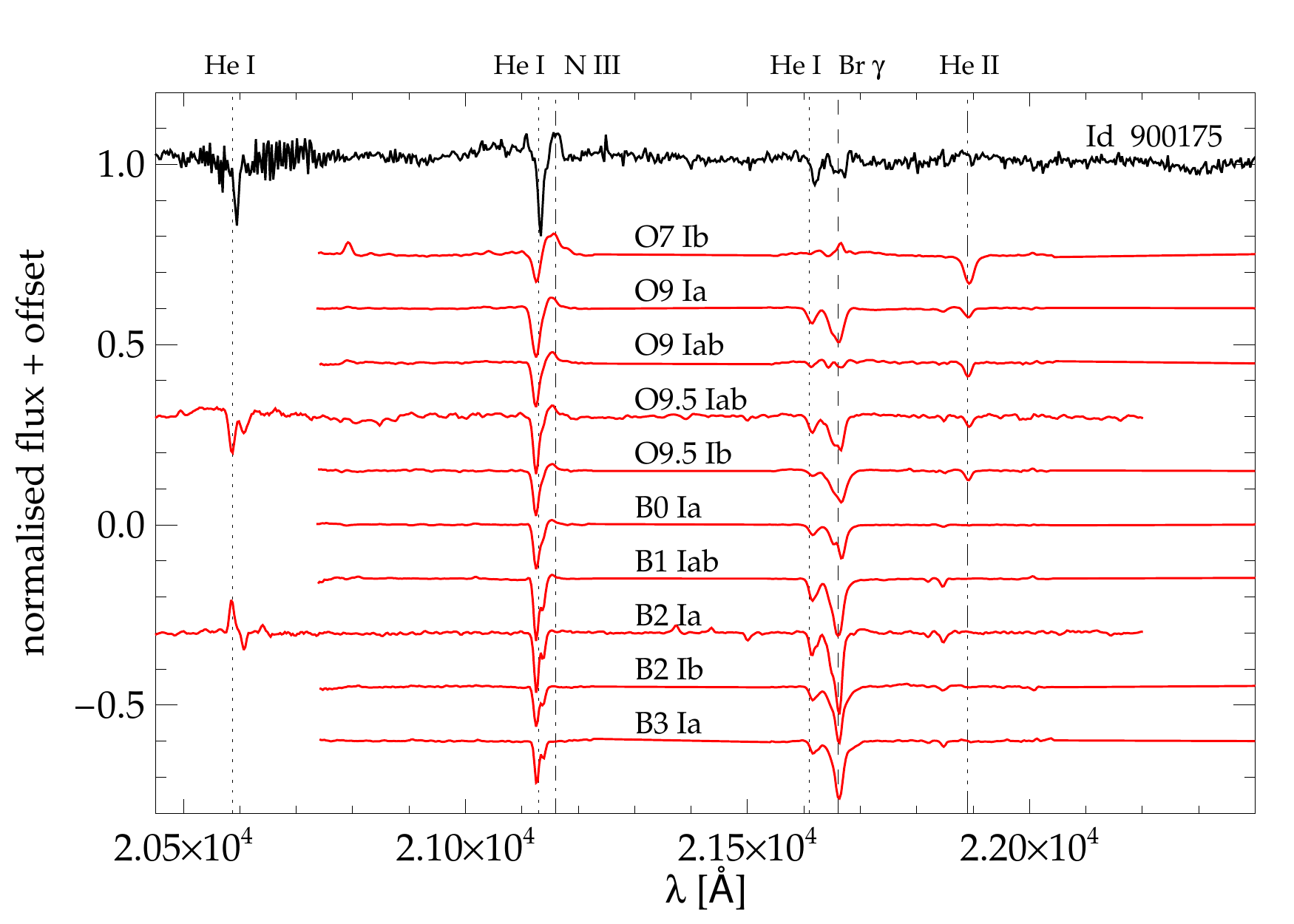}
 \caption{Spectrum of Id 900175 (black) in comparison to the \citet{2005ApJS..161..154H} spectra of O/B type supergiant stars (red), brought to the spectral resolution of our data. }
 \label{fig:id900175}
\end{figure}

Id 900175 is the brightest OB star in our sample with the highest S/N. It has a P Cygni profile at 2.113\,\micron. The shape of this profile suggests the luminosity class I rather than II or higher, and a O8 to O9.5 spectral type. Also the He\,{\sc i}\space \normalsize 2.161\,\micron\space and Br\,$\gamma$ absorption support this classification. We show several spectra of supergiant stars from \cite{2005ApJS..161..154H} in comparison to Id 900175 in Fig.~\ref{fig:id900175}. From this spectral classification and the statistical relations by \cite{2020RAA....20..139M}, we derived a more narrow \teff\space range of 28900--36700\,K for Id 900175.

\begin{figure}
 \centering 
 \includegraphics[width=0.99\columnwidth]{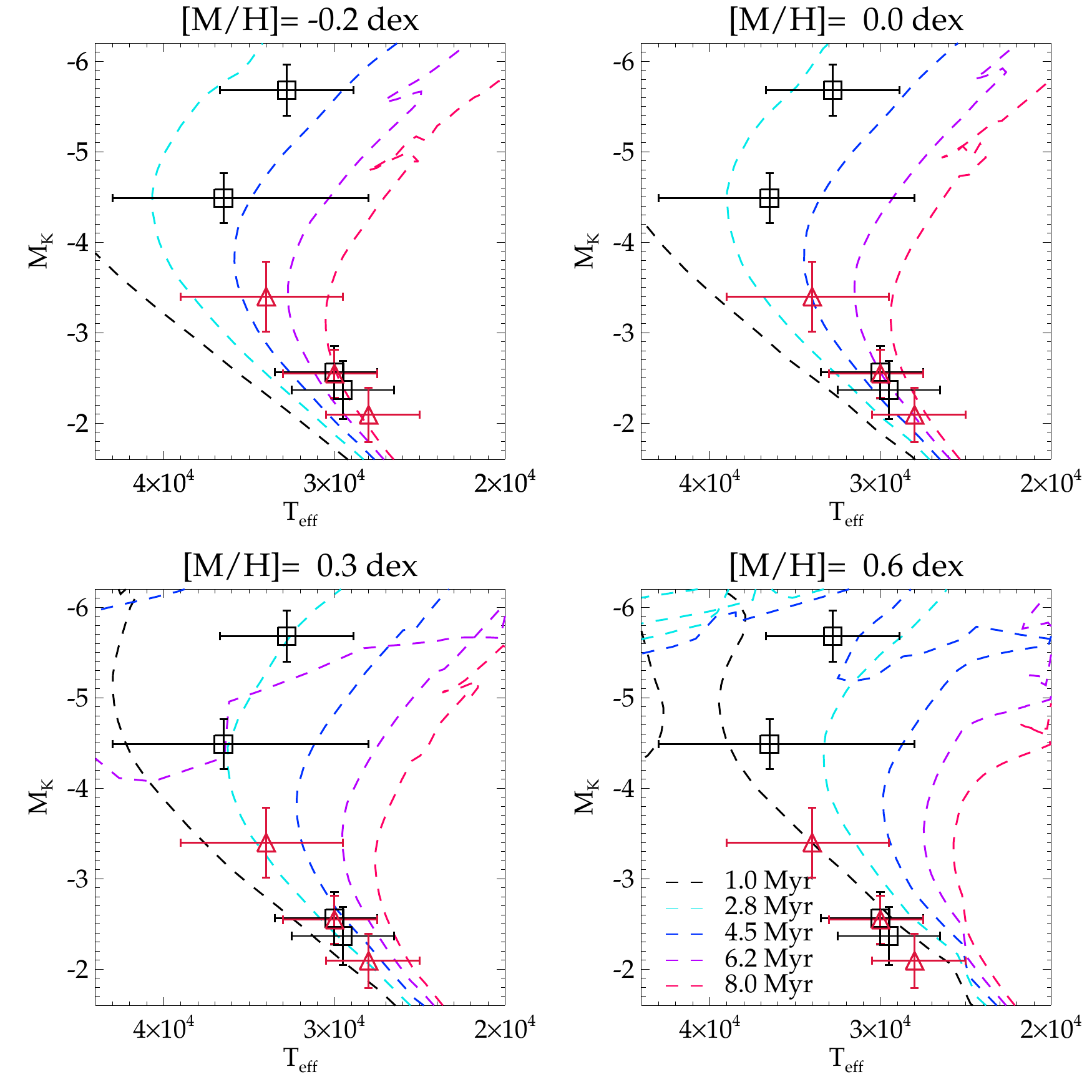}
 \caption{Hertzsprung-Russel diagram of OB star candidates in the East (black squares) and West (red triangles) fields. Coloured dashed lines denote PARSEC isochrones with varying ages (see legend in lower right panel), the four panels different values for \mh, as noted on the top.}
 \label{fig:tefflum}
\end{figure}

We used the same PARSEC isochrones and our estimates of \teff\space and the absolute stellar magnitude $M_K$ to test whether we can exclude any of the isochrones for the two fields. We assumed that all the young stars per field formed in the same star formation event, and that they have a distance of 8.178\,kpc \citep{2019A&A...625L..10G} with a distance uncertainty of 150\,pc. Our OB star sample and several isochrones are also shown in a Hertzsprung-Russell diagram in Fig.~\ref{fig:tefflum}. We computed the weighted sum of squared residuals \citep{1987mem..book.....F} for each star and isochrone point, used its minimum per star and isochrone, and summed over the five East and three West stars. 
We find that both fields prefer similar ages ($\sim$2.5--5.5\,Myr and 3--6.5\,Myr) and supersolar \mh\space ($\sim$0.2--0.5\,dex). Populations with 1--2.5\,Myr age and \mh=--0.2\,dex, or 7--8\,Myr and \mh=0.6\,dex are excluded ($\gtrsim$2$\sigma$ confidence). However, we note that our results depend on the reference intrinsic colours that we use to derive $M_K$ and \teff, and thus the age and \mh\space may be biased. 

\begin{figure}
 \centering 
 \includegraphics[width=0.99\columnwidth]{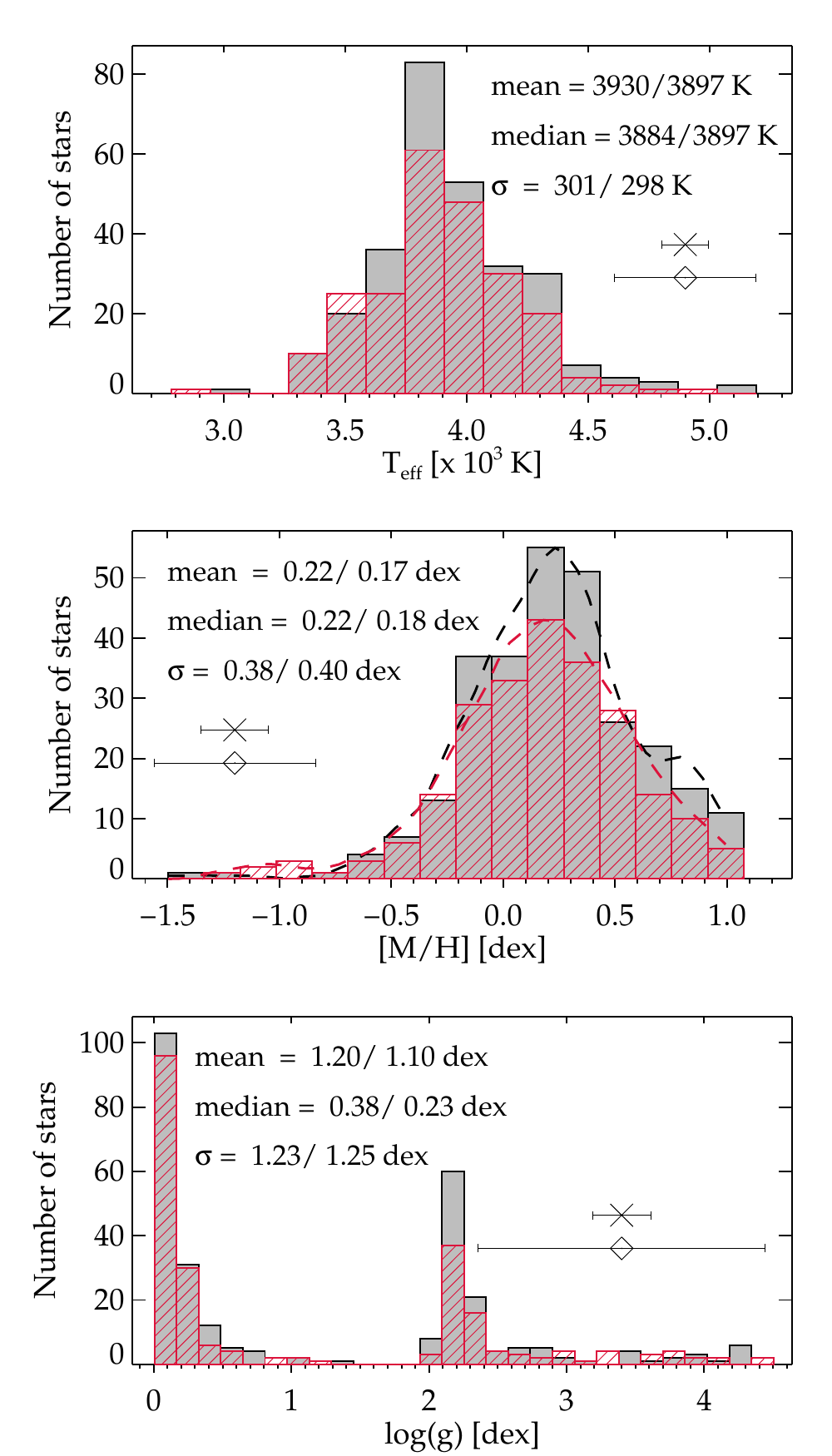}
 \caption{Histograms of stellar parameter measurements of red giant stars for the East (black) and West (red) fields. From top to bottom: effective temperature $\teff$, metallicity \mh, surface gravity $\logg$. On each panel we list the mean, median and standard deviation of the distributions (first East, then West field). The error bars with x-symbol represent the mean statistical uncertainties, the diamond symbol the mean total uncertainties, which are similar in both fields. Dashed lines denote the Kernel density estimation of the \mh\space distributions.}
 \label{fig:meanhist}
\end{figure}


\subsection{Stellar Parameter Distributions of Red Giant Stars}

Our final stellar parameter distributions for \goodne\space red giant stars in the East and \goodsw\space red giant stars in the West are shown in Fig.~\ref{fig:meanhist}. We only show stars that are located at the distances of the Galactic centre based on their $H-K_S$ colours. The distributions of \teff\space (top), \mh\space (middle), and \logg\space (bottom) are similar among the two fields, with similar mean, median, and standard deviations, as labelled in Fig.~\ref{fig:meanhist}. The East field has slightly larger values of \mh\space and \logg, but the differences are much smaller than the mean measurement uncertainties. 

We obtain higher values for the mean \teff \space ($\sim$3900\,K), \logg \space ($\sim$1.15\,dex), and lower values of \mh\space ($\sim$0.2\,dex)
in comparison to the central 4\,pc$^2$ \citep[\teff=3600\,K, \mh=0.26\,dex, \logg=0.5\,dex]{2017MNRAS.464..194F}, and the inner r\,\textless 5\,pc \citep[3500\,K, 0.34\,dex, 0.34\,dex]{2020MNRAS.494..396F}. The higher values for \teff\space and \logg\space are probably caused by our deeper observations, which include a larger fraction of relatively warmer and less massive stars. We look more into the difference of \mh\space in the following section.
Our stellar parameter measurements are available as supplementary material.

\subsection{Comparison to NSC and NSD Metallicities}
\label{sec:profile}
\begin{figure*}
 \centering 
 \includegraphics[width=\textwidth]{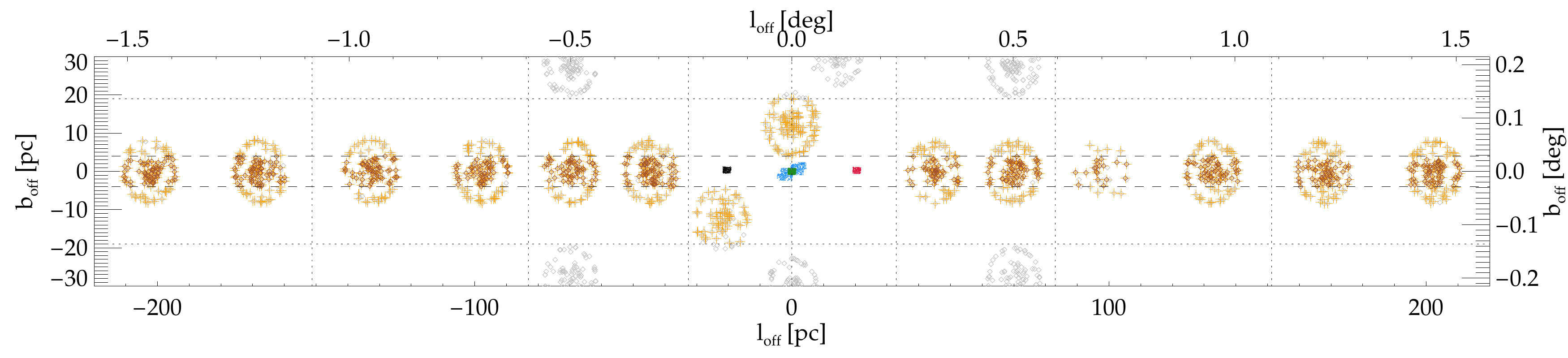}
 \caption{Spatial coverage of the stars used to measure the metallicity profile in Fig.~\ref{fig:mhvsl}. Black and red data points denote the East and West data, green data from \citet{2017MNRAS.464..194F}, blue from \citet{2020MNRAS.494..396F}. Not all data from \citet[grey]{2021A&A...649A..83F} were used, only stars shown in orange (--19\,pc\textless b$_\text{off}$\textless 19\,pc, dotted horizontal line) and brown (--4\,pc\textless b$_\text{off}$\textless 4\,pc, dashed horizontal line). Vertical dotted lines mark the binning of l$_\text{off}$.} 
 \label{fig:selectedfritz}
\end{figure*}

We compared our metallicity measurements of stars in the transition region of the NSC to the NSD with the central NSC data of \cite{2017MNRAS.464..194F,2020MNRAS.494..396F}, and with the NSD data of \cite{2021A&A...649A..83F} along the Galactic plane. The spatial coverage of the different studies is shown in Fig.~\ref{fig:selectedfritz}. 

All these data are based on KMOS observations, however, \cite{2021A&A...649A..83F} derived an empirical [Fe/H] calibration using $EW_\text{CO}$ and $EW_\text{Na}$, while all the NSC data and this work used full-spectral fitting to measure [M/H], from which the wavelength regions around $EW_\text{Na}$ and $EW_\text{CO}$ are excluded. To test whether the results of [M/H] are comparable to [Fe/H] as measured by \cite{2021A&A...649A..83F}, we applied the \cite{2021A&A...649A..83F} [Fe/H]$_\text{F21}$-spectral index relation to our data and the data of \cite{2017MNRAS.464..194F,2020MNRAS.494..396F}. 
 We decided to limit the data to the ranges 0\,$\angstrom$\textless $EW_\text{CO}$ \textless 25\,$\angstrom$, and 0\,$\angstrom$\textless $EW_\text{Na}$ \textless 7\,$\angstrom$, since this is where the majority of the empirical spectra lie that are used to calibrate the [Fe/H]$_\text{F21}$-spectral index relation.
 The results largely agree \citep[see also ][]{2021A&A...650A.191S}, with the median $\lvert$[M/H]--[Fe/H]$_\text{F21}\rvert$ in the range 0.0013 to 0.02\,dex for \cite{2017MNRAS.464..194F,2020MNRAS.494..396F} and the data of this study. 
 The 25th percentiles differ by 0.004 to 0.07\,dex, and the 75th percentiles by -0.014 to 0.003\,dex. 
 The standard deviation of [M/H]--[Fe/H]$_\text{F21}$ is about 0.5\,dex, which is less than the combined uncertainties. 
 We note that [M/H]\textgreater[Fe/H]$_\text{F21}$ at $EW_\text{Na}\lesssim$1.5\,$\angstrom$, and the scatter of [M/H]--[Fe/H]$_\text{F21}$ increases with decreasing $EW_\text{CO}$. 
 Nevertheless, overall there is no systematic bias between [M/H] and [Fe/H]$_\text{F21}$, and we can compare the different metallicity measurements. 

\begin{figure}
 \includegraphics[width=0.98\columnwidth]{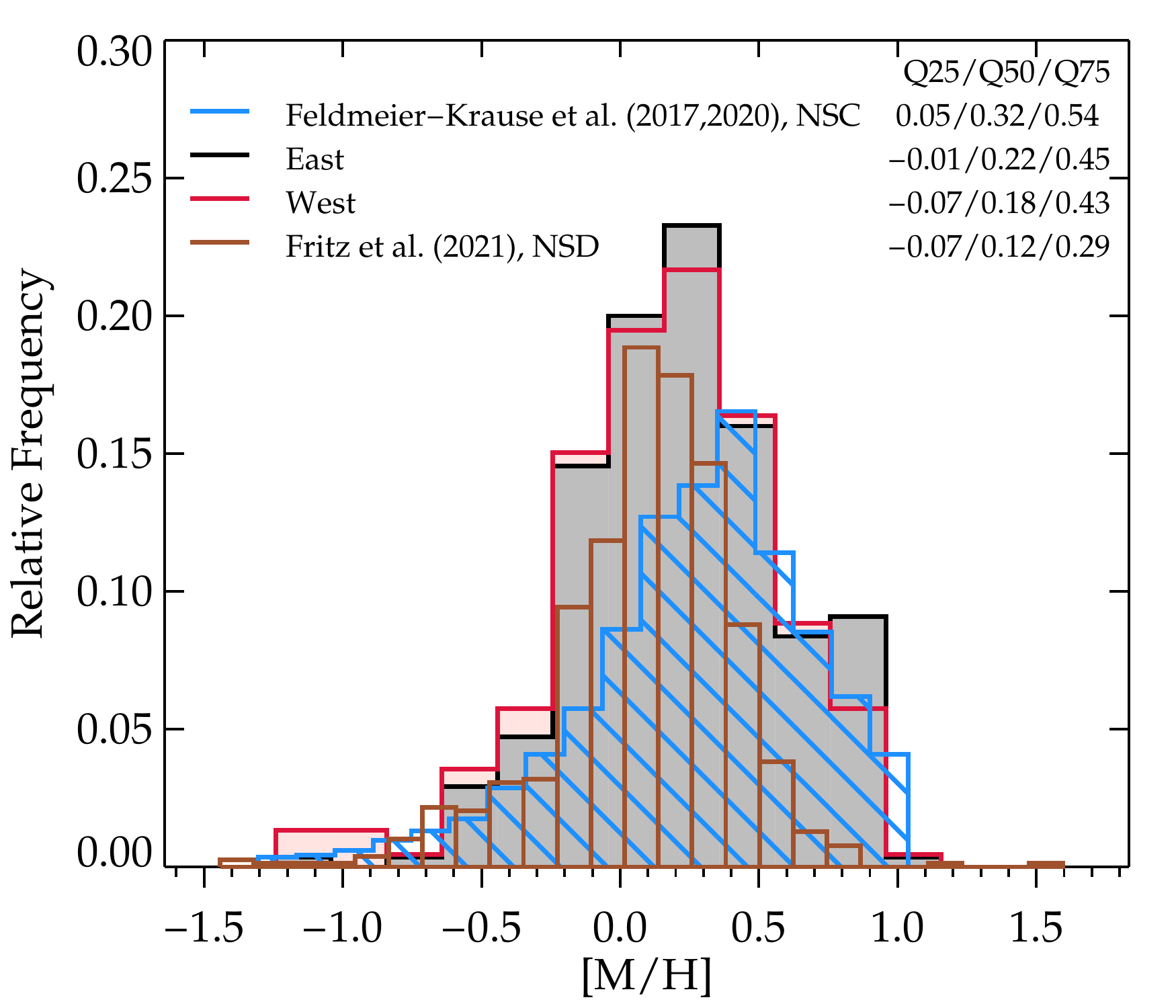}
 \caption{The histograms of \mh\space for the East (black), and West (red) fields are similar. NSC data of \citet[blue]{2017MNRAS.464..194F,2020MNRAS.494..396F} is shifted to higher \mh, NSD data of \citet[brown, $\lvert$b$_\text{off}\rvert$\textless 4 pc]{2021A&A...649A..83F}, which shows [Fe/H]$_\text{F21}$, to lower values, see also quartiles for 25th, 50th, and 75th percentiles shown in upper right. }
 \label{fig:mh0}
\end{figure}

\begin{figure}
 \centering 
 \includegraphics[width=0.99\columnwidth]{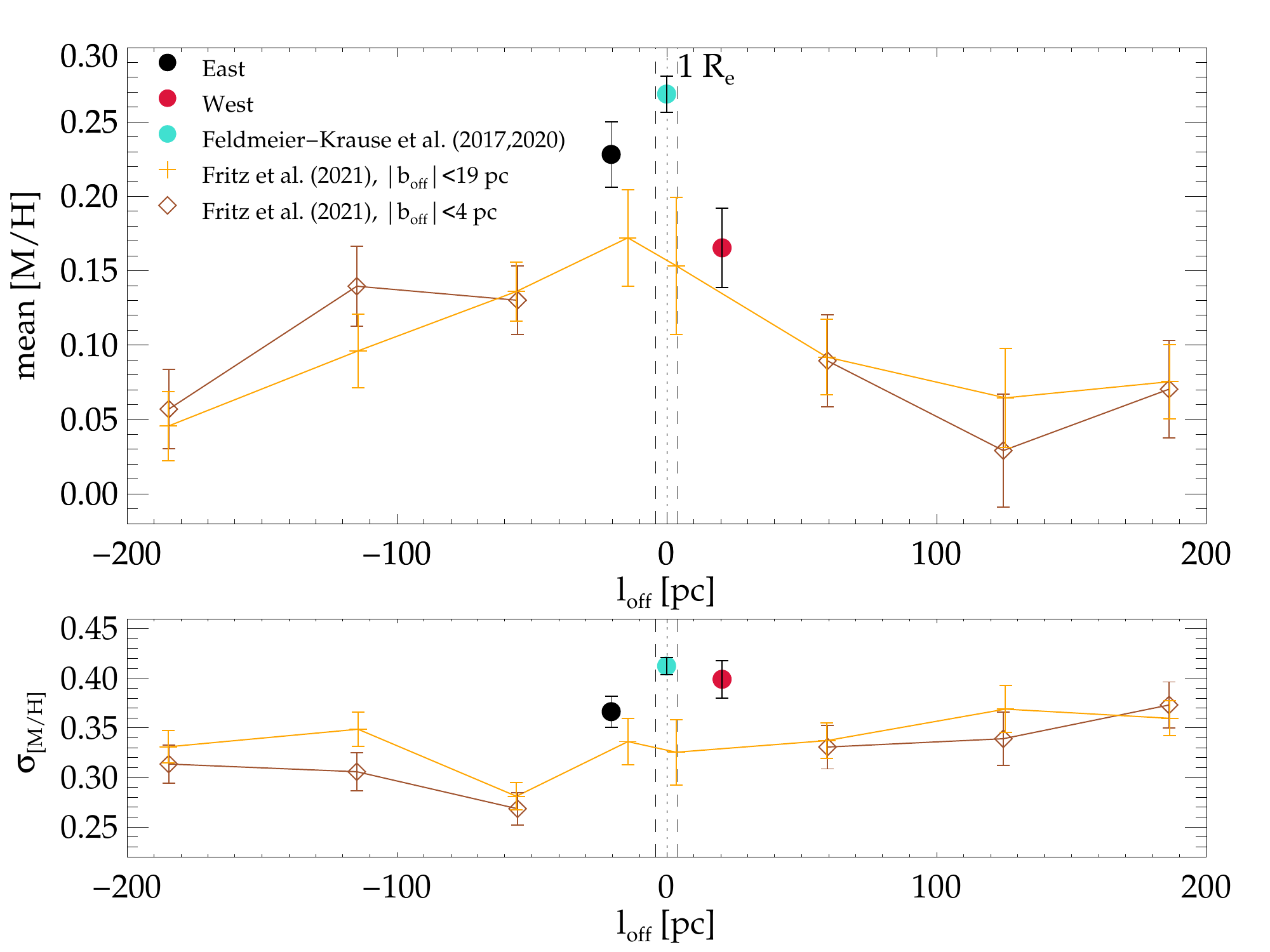}
 \caption{Mean metallicity profile (top) and standard deviation $\sigma$ as function of offset Galactic longitude l$_\text{off}$ from Sgr A*. Different colours denote different data sets and binnings as shown in Fig.~ \ref{fig:selectedfritz}. The standard error of the mean is $\sigma$ divided by the square root of the number of stars per bin $N$, the standard error of the standard deviation is $\sigma$/$\sqrt{2\cdot N-2}$. }
 \label{fig:mhvsl}
\end{figure}
\begin{figure}
 \centering 
 \includegraphics[width=0.99\columnwidth]{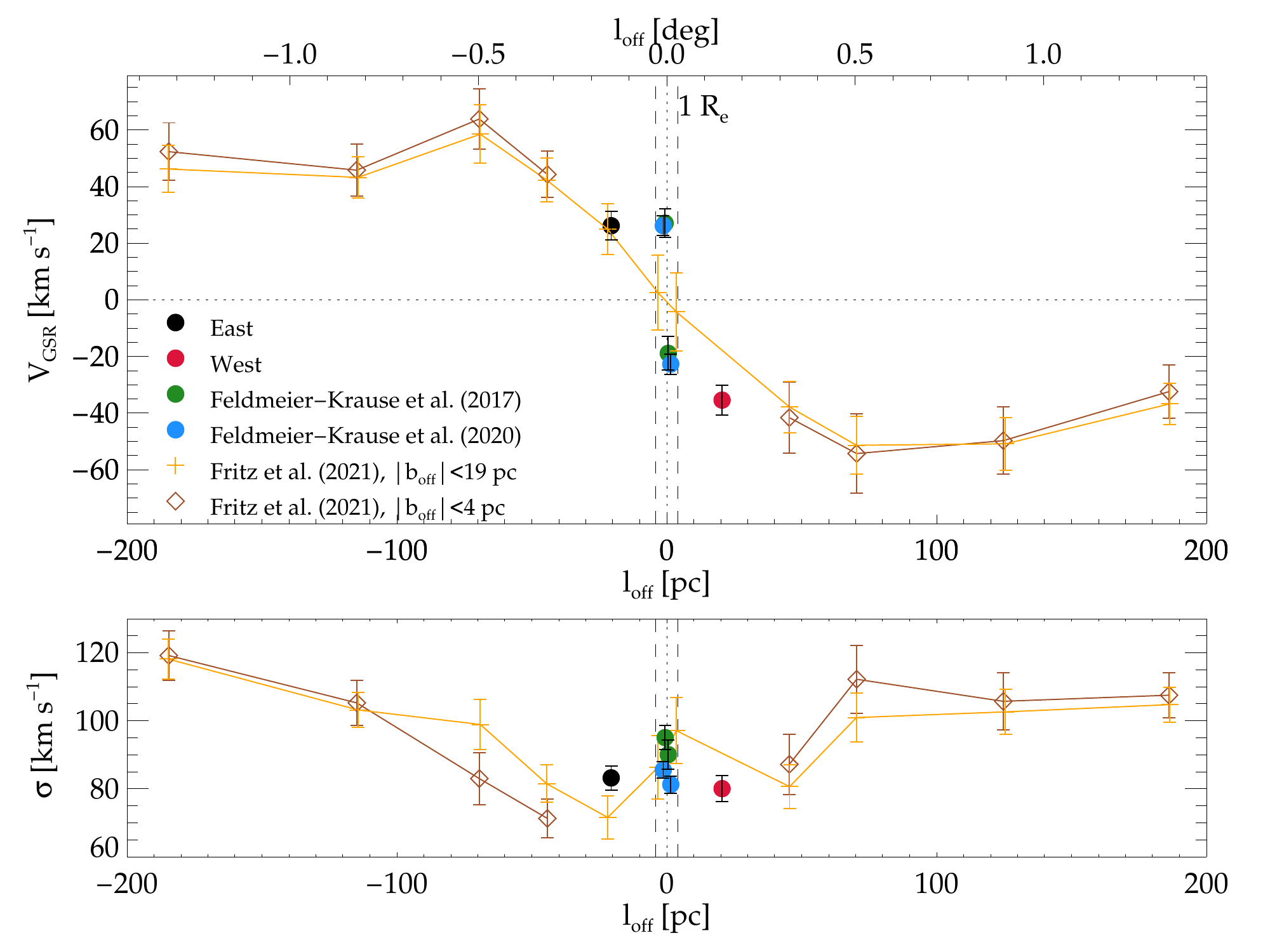}
 \caption{Mean velocity profile (top) and velocity dispersion $\sigma$ (computed following \citealt{1993ASPC...50..357P}) as function of offset Galactic longitude l$_\text{off}$ from Sgr A*. }
 \label{fig:rvvsl}
\end{figure}

 We show the metallicity distributions of the samples in Fig.~\ref{fig:mh0}, and list Q25, Q50, and Q75, which denote the 25th, 50th, and 75th percentiles of the metallicity distribution, in the upper right. The NSC data \citep{2017MNRAS.464..194F,2020MNRAS.494..396F} are shown in blue, our East and West fields in black and red outline, and the NSD data \citep{2021A&A...649A..83F} in brown. For the latter we show [Fe/H]$_\text{F21}$ instead of [M/H], only stars that are highlighted in brown colour in Fig.~\ref{fig:selectedfritz} within $-4$\,pc\textless b$_\text{off}$\textless 4\,pc, and with statistical uncertainty $\sigma_{[Fe/H]}\leq$0.35\,dex. This is slightly higher than the minimum statistical uncertainty we chose for full spectral fitting, but using the same value (i.e. $\sigma_{[Fe/H]}\leq$0.25\,dex) biases the standard deviation of the distribution to much lower values.
 The entire distributions and thus the percentiles shift with increasing $\lvert$l$_\text{off}\rvert$, which is the distance from Sgr A* along the Galactic longitude, to lower \mh\space values. Q25 shifts by only 0.12\,dex, Q50 by 0.2, and Q75 by 0.25\,dex from the NSC to the NSD. Thus, the difference of the \mh-distributions of NSC and NSD is larger at high \mh. 
The East and West fields are usually in between these values, only Q25 of the West field equals Q25 of the NSD sample.

 In Fig.~\ref{fig:mhvsl} we show the mean metallicity (top panel) and standard deviation $\sigma_\text{\mh}$ (bottom panel) as a function of Galactic longitude offset from Sgr A* (l$_\text{off}$). 
 NSD data are binned as shown in Fig.~\ref{fig:selectedfritz} in two different ranges of offset Galactic latitude b$_\text{off}$, 8\,pc and 38\,pc wide. The orange data points cover the wider latitude range and show less scatter, but the results are similar. The outer NSD has a mean metallicity $\sim$0.05\,dex, increasing to $\sim$0.15\,dex towards the centre. The mean metallicity of the central NSC is even higher ($\gtrsim$0.25\,dex). In our East and West fields at a distance of 20\,pc, which is more than 4\,\re\space from the NSC, the mean metallicity is higher than in the NSD, but the measurements agree within 1\,$\sigma$. The standard deviation of the metallicity measurement $\sigma_\text{\mh}$ is $\geq$0.05\,dex higher than in the NSD. This is not caused by the different metallicity measurement methods, as we obtain similar results when we compare the standard deviation of [Fe/H]$_\text{F21}$ for the NSC transition data. However, when we restrict the NSC data to the same range of $K_{S,0}$ as the NSD data (7-9.5\,mag), the value of $\sigma_\text{[M/H]}$ decreases, to almost 1$\sigma$ agreement with the NSD data. It is possible that the NSD data have the lower value of $\sigma_\text{[M/H]}$ compared to the NSC because it samples only a rather narrow range of $K_{S,0}$. The mean metallicity, however, does not change significantly when restricting NSC data to the magnitude range of the NSD data.

We tested if the different completeness levels of \cite{2021A&A...649A..83F}, \cite{2020MNRAS.494..396F}, and our data set have any effect on the resulting metallicity distribution. We used simulated stellar populations of PARSEC version 1.2S \citep{2014MNRAS.445.4287T} with a \cite{2001MNRAS.322..231K} initial mass function. We downloaded a 1 Gyr old, \mh=0.2\,dex, a 3 Gyr old, \mh=0.2\,dex, and several 8 Gyr old stellar populations with \mh\space ranging from -1\,dex to 0.5\,dex, each with a mass of 10$^6$\,M$_{\sun}$. From this set of simulated stellar populations we constructed mixed stellar populations that resemble the Milky Way NSC and NSD star formation history and metallicity distribution. 

We did this by drawing random stars from each stellar population. We considered that $\sim$80\% of the NSC stellar mass is $\sim$8\,Gyr old, and $\sim$15\% is 3\,Gyr old \citep{2020A&A...641A.102S}, neglecting any younger stars. For the NSD we assumed that $\sim$90\% of stars are 8\,Gyr and $\sim$5\% are 1\,Gyr old \citep{2020NatAs...4..377N}, again neglecting younger stars. To mimic the metallicity distributions, we added errors drawn from a normal distribution with width 0.3\,dex to the individual stellar metallicities. Then, we drew random stars from the populations, and the fractions of the individual populations were chosen such that the metallicity distribution of the final simulated population resembles the quantiles of the respective data sets (see Figure \ref{fig:mh0}). On this simulated mixed population we applied the completeness cuts of the different data sets. This means we selected only a certain fraction of stars, depending on the completeness in their magnitude range. From this incomplete representation of the simulated mixed population we computed quantiles, the mean and standard deviation of the metallicity distribution. We repeated this process in 1000 runs, and compared the resulting \mh\space distributions with the complete simulated mixed population. We found small shifts to lower mean metallicity by \textless0.005\,dex for the NSC and \textless0.04\,dex for the NSD. These shifts are too small to explain the different \mh\space found in the NSC and NSD. The standard deviation of the incomplete populations is even slightly larger compared to the complete population, by 0.007\,dex for our data set, and 0.06\,dex and 0.08\,dex for \cite{2020MNRAS.494..396F} and \cite{2021A&A...649A..83F}. We conclude that the different completeness levels of the data have a negligible effect on the metallicity distributions.

We measured \mh\space gradients from the mean value of the central NSC data to our East and West fields separately and folded as function of the $\lvert$l$_\text{off}\rvert$, the results are listed in Table \ref{tab:gradmh}. For comparison, we measured the [Fe/H]$_\text{F21}$ gradients, considering only the mean [Fe/H]$_\text{F21}$ values of NSD data. Using a wider latitude range does not change these beyond the uncertainties. Our gradients show that the high \mh\space of the NSC decreases steeply with radius towards 20\,pc, with a gradient --0.032\,dex/10\,pc, and is by a factor $\sim$8 larger than the metallicity gradient of the NSD (--0.004\,dex/10\,pc). If we use the gradient of the NSD, we obtain a mean \mh\space of 0.15$\pm$0.03\,dex at 0\,pc, which is 3$\sigma$ lower than the actual 0.27\,dex. We conclude that the steep \mh\space increase towards the centre cannot be explained by a linear continuation of the NSD, but is rather caused by a different stellar population, belonging to the NSC.

\begin{table}
 \centering
\caption{\mh\space gradient along Galactic offset longitude l$_\text{off}$}
 \label{tab:gradmh}
\begin{tabular}{@{}lclll@{}}
\hline
&$\lvert b_\text{off}\rvert$& East&West &folded $\lvert$l$_\text{off}\rvert$\\
&pc&	dex/10 pc&dex/10 pc&dex/10 pc\\
\hline
NSC to E/W & \textless 2.4 & 0.020 $\pm$ 0.012 &--0.050 $\pm$ 0.014 & --0.032 $\pm$ 0.010 \\
NSD & \textless 4 & 0.006 $\pm$ 0.003 &--0.002 $\pm$ 0.004 & --0.004 $\pm$ 0.002 \\
 \hline 
 \end{tabular}
\end{table}

We also show the stellar kinematic profiles (computed as in \citealt{1993ASPC...50..357P}) in Fig.~\ref{fig:rvvsl} in a similar way. All velocities were converted to the Galactocentric velocity frame \citep[$V_\text{apex}$=232.3\,\kms, l$_\text{apex}$=87.8\degr, b$_\text{apex}$=1.7\degr]{1991rc3..book.....D}. 
We used a finer binning than in Fig.~\ref{fig:mhvsl} in the centre to see the shape of the NSC rotation curve. 
While the NSC data points at negative l$_\text{off}$, i.e. in the Galactic East, are offset by \textgreater 1\,$\sigma$ from the NSD data, the velocities of the two transition regions at $\lvert$l$_\text{off}\rvert\sim$20\,pc are $\lesssim$ 1\,$\sigma$ offset to the NSD data. The velocity dispersion of the NSC data is within the scatter of the NSD data points. 
The velocity dispersion of the stars in the transition region is similar to the outer NSC, $\sim$82\,\kms, the mean velocity indicates rotation that is in agreement with the NSD.


\section{Discussion}
\label{sec:sec5}

\subsection{The Origin of Young Stars}
We detected several young star candidates located $\sim$20\,pc from Sgr A*, along the Galactic plane. We found that 4-6\,Myr old isochrones fit the data best, though in principle ages in the range 1-10\,Myr are possible. Here we discuss the potential origin of these stars.

Young, isolated stars in the Galactic centre have been detected in the past \citep[e.g.][]{1999ApJ...510..747C,2010ApJ...710..706M,2010ApJ...725..188M}. It was suggested that these stars may have formed in isolation or in small groups \citep{2011MNRAS.417..114D}. The unique conditions in the Galactic centre (e.g. high gas density, high temperature, strong tidal field and magnetic field) 
may indeed require a different form of star formation than in the Galactic disc, where massive star formation happens in clusters. \cite{2015MNRAS.446..842D} suggested star formation associated to bow shocks or pressure-driven flows for some of their massive young field stars.

On the other hand, the stars may have formed in clusters and were ejected via SN explosions \citep{1961BAN....15..265B}, in which they received a kick, or in dynamical interactions \citep{1967BOTT....4...86P}. The cluster candidates for their origin are the three young clusters in the Galactic centre: 
The central parsec cluster, Arches and Quintuplet. These clusters have ages that are similar to our OB stars candidates \citep{1999ApJ...514..202F,2004ApJ...611L.105N,2006ApJ...643.1011P}. 
While the central parsec cluster is located around Sgr~A*, Arches and Quintuplet are located in the Galactic East. While Arches has a projected distance of $\sim$9.5\,pc to the North-North-East of our East field, Quintuplet is located $\sim$10\,pc to the East-South-East of our East field. Arches appears to be the youngest cluster \citep[2-3 Myr,][]{2018A&A...617A..65C}, which makes SN explosions unlikely, but they may be possible for the central parsec cluster and Quintuplet. 
If a star in a binary system evolves to a SN, it can accelerate the companion to $\gtrsim$100\,\kms, thus displacing the star by up to 100\,pc within 1\,Myr \citep{2021A&A...649A..43C}, but in random directions. We detected nine stars in only a small region ($\sim$8.5\,pc$^2$), which makes this mechanism unlikely to be responsible for all the stars.

Also 3-body interactions of a single star with a binary or multiple system are able to accelerate stars. \cite{2011MNRAS.410..304G} used $N$-body simulations and found that stars in the mass ranges 12--20\,M$_{\sun}$, where most of the stars in our data are, are accelerated 
in \textgreater50\% of encounters with a massive binary, typically to $\sim$90\,\kms. For more massive stars (38--60\,M$_{\sun}$), the average recoil velocity is $\lesssim$50\,\kms, though in some cases velocities up to 100\,\kms\space can be reached.

Another dynamical interaction is tidal stripping of stars from the clusters. For this process, Quintuplet seems to be the best option: Its proper motions are along the Galactic plane, similar to Arches, moving towards the East \citep{2008ApJ...675.1278S,2014ApJ...789..115S}. Within its lifetime, it passed at least once at a projected distance of 2\,pc South of both the East field and West field, assuming no vertical motion \citep{2014ApJ...789..115S}. \cite{2014A&A...566A...6H} simulated the dynamical evolution of Quintuplet and Arches and found that they develop tidal tails of drifted sources, that extend along the cluster orbits. The stars in our fields may be just part of these tails. In that case, they would have proper motions moving in the same direction as Quintuplet, but trailing behind due to lower velocities. 
It is also possible that the stars formed in another cluster that has dissolved already. Due to a lower density than e.g. Quintuplet, clusters may become too dispersed to be detected in only $\sim$5\,Myr \citep{2001ApJ...546L.101P,2002ApJ...565..265P}. In this case, the stars would have proper motions pointing to the same direction \citep{2019A&A...632A.116S}. 

We do not have proper motions to test these scenarios, only line-of-sight velocities, which are reported in Table \ref{tab:obtab}. The current line-of-sight velocity of Quintuplet is 102$\pm$2\,\kms\space \citep{2009A&A...494.1137L}. Since it is unclear at what line-of-sight distance Quintuplet is, its exact orbit and its line-of-sight velocity at the locations close to our two fields are unknown. 
The velocity dispersion of Quintuplet \citep[17\,\kms,][]{2009A&A...494.1137L} is comparable to the three OB stars in the West field (21$\pm$11\,\kms), but lower than for the five stars in the East field ($\sigma$=53$\pm$24\,\kms). This indicates that the stars in the West field may be entirely part of a tidal tail associated to Quintuplet, while stars in the East tail may originate from a combination of the aforementioned processes.

\subsection{Distinct Formation Histories of the NSC and NSD}
The NSC is located within the NSD, which suggests that there is a connection of these structures. 
Yet, there are several differences, which indicate that NSD and NSC are distinct components. 

We compared the stellar metallicities and kinematics of red giant stars in the NSD, the NSC, and in the transition region at a distance of 20\,pc (\textgreater 4\,\re) from the NSC. We found that the mean stellar metallicity is decreasing outwards, from the NSC to the NSD, and the transition region lies in between. The volume densities of the NSC and NSD at the transition region have the same order of magnitude, while the density of the NSC at 1\,\re\space is higher by a factor $\sim$15, and in the very centre by a factor $\sim$100 compared to the NSD \citep{ 2022MNRAS.512.1857S}. The stars in the transition regions are not dominated by one of the two structures, but rather a roughly equal mix of stars belonging to the NSC and NSD.
The [M/H] gradient from the central NSC to the transition region (r=20\,pc) is much steeper than the gradient of the NSD at r=50-200\,pc. The steepness in the central 20\,pc is probably not due to a steep change of \mh\space in the NSC or NSD, but rather due to the increasing number of NSD stars in the transition region, which tend to have a lower \mh\space than NSC stars. The standard deviation of \mh\space in the transition region is higher than in the NSD, as it is a mix of NSC and NSD stars.

NSD and NSC are likely connected via gas inflow, though
it is unclear how efficiently gas is brought in from the NSD to the NSC and forms stars there. The inflow rate caused by supernova feedback towards the inner 50\,pc fluctuates in time, with a time average of $\sim$0.03\,M$_{\sun}$\,yr$^{-1}$ \citep{2020MNRAS.499.4455T}. Other potential drivers of a nuclear gas inflow may be magnetic fields, a nuclear bar, or external perturbers (e.g. star clusters, satellites). If gas is enriched in the NSD and then flows inwards to the NSC and forms stars there, it would be a natural consequence that NSC stars have higher \mh.

However, the star formation histories of the Milky Way's NSC and NSD have several differences \citep{2020NatAs...4..377N,2020A&A...641A.102S,2021ApJ...920...97N}: 
While $\sim$5 per cent of the stars in the NSD are $\sim$1\,Gyr old, there are no signs of star forming activity for that time in the NSC. The NSC, however, formed $\sim$15 per cent of its stars 3\,Gyr ago, but there appear to be no such intermediate-age stars in the NSD. It seems that star formation activity followed by supernova feedback in the NSD is not the only way to trigger star formation in the NSC.

The higher metallicity of the NSC compared to the NSD, and their different star formation histories indicate that the NSC is not only a continuation of the NSD but has a distinct formation history. 
Extragalactic studies suggest that one structure can exist without the other. 
Many barred galaxies with NSD lack a hot spheroidal central component, such as a NSC \citep{2020A&A...643A..65B}. 
 While it is possible that the NSCs are too small and faint to be detected, it is also possible that the NSDs exist without a NSC. On the other hand, there are galaxies with a NSC but no NSD, such as NGC 300. 
Thus, NSCs and NSDs can co-exist in the same galaxy, but they can also exist independently.
This suggests that either of these structures can form independently, though it may also be possible that one or the other can be destroyed while the other structure perseveres.

\section{Conclusions}
 \label{sec:sec6}
 
 We analysed stellar spectra in two fields along the Galactic plane, $\sim$20\,pc East and West from Sgr A*. Each field covers $\sim$4.3\,pc$^2$. 

Among the \textgreater200 stars per field, we detected five OB star candidates in the East field, and three OB and one Wolf Rayet star candidates in the West field. We used the OB stars' photometry and theoretical isochrones to estimate the stellar masses, \teff, and spectral types. The positions of the stars in the Hertzsprung Russel diagram show that super-solar, 4-6\,Myr old isochrones fit best, though in principle ages in the range 1-10\,Myr are possible. We discuss the origin of these stars and suggest that they come from a combination of dynamical processes, including tidal stripping from the Quintuplet cluster. 

The majority of stars (281 and 228 in East and West) are red giants, and we measured their stellar metallicities. We compared the metallicity distribution to the NSC and NSD metallicity distributions \citep{2017MNRAS.464..194F,2020MNRAS.494..396F,2021A&A...649A..83F} and found a continuous decrease of metallicity with increasing distance to Sgr A* along the Galactic plane. Not only the mean metallicity, but also the 25th, 50th, and 75th quantiles of the metallicity distributions shift to lower [M/H] with increasing Galactocentric distance. The volume densities of NSC and NSD are roughly equal in the transition region, and we have a mix of the two components. 
From the NSC dominated centre to the transition region (r=0-20\,pc), the mean [M/H] gradient is 8 times steeper than the gradient of the NSD (r=50-200\,pc) along the Galactic longitude. 
The \mh\space distribution of the NSC is systematically higher than in the NSD, and we suggest that the two components have 
 distinct stellar populations and formation histories.

 \section*{Acknowledgments}
We thank the ESO staff who helped us to prepare our observations and obtain the data. We thank Lodovico Coccato and Yves Jung for advice and assistance in the data reduction process. 
We are grateful to Nadine Neumayer for useful comments and suggestions. We also thank the referee for constructive comments
and suggestions.
Based on observations collected at the European Organisation for Astronomical Research in the Southern Hemisphere, Chile (60.A-9450(A), 093.B-0368(A), 195.B-0283, 0101.B-0355(A)).
This research has made use of the SIMBAD database, operated at CDS, Strasbourg, France. This research has made use of NASA's Astrophysics Data System. This research made use of ds9, a tool for data visualization supported by the Chandra X-ray Science Center (CXC) and the High Energy Astrophysics Science Archive Center (HEASARC) with support from the JWST Mission office at the Space Telescope Science Institute for 3D visualization. This research made use of Montage. It is funded by the National Science Foundation under Grant Number ACI-1440620, and was previously funded by the National Aeronautics and Space Administration's Earth Science Technology Office, Computation Technologies Project, under Cooperative Agreement Number NCC5-626 between NASA and the California Institute of Technology.

 \section*{Data Availability}

The data underlying this article will be shared on reasonable request to the corresponding author.

\footnotesize{
\bibliography{bibs_lt}
}
\normalsize
\appendix

\section{Systematic uncertainty estimation of stellar parameters}
\label{sec:xsl}

In order to estimate the systematic uncertainty of our stellar parameter fits, we fit stellar spectra with known stellar parameters in similar ranges as our data. We applied the same method as for our data, and compared our results with the reference literature values.

\begin{figure*}
 \centering 
\includegraphics[width=\textwidth]{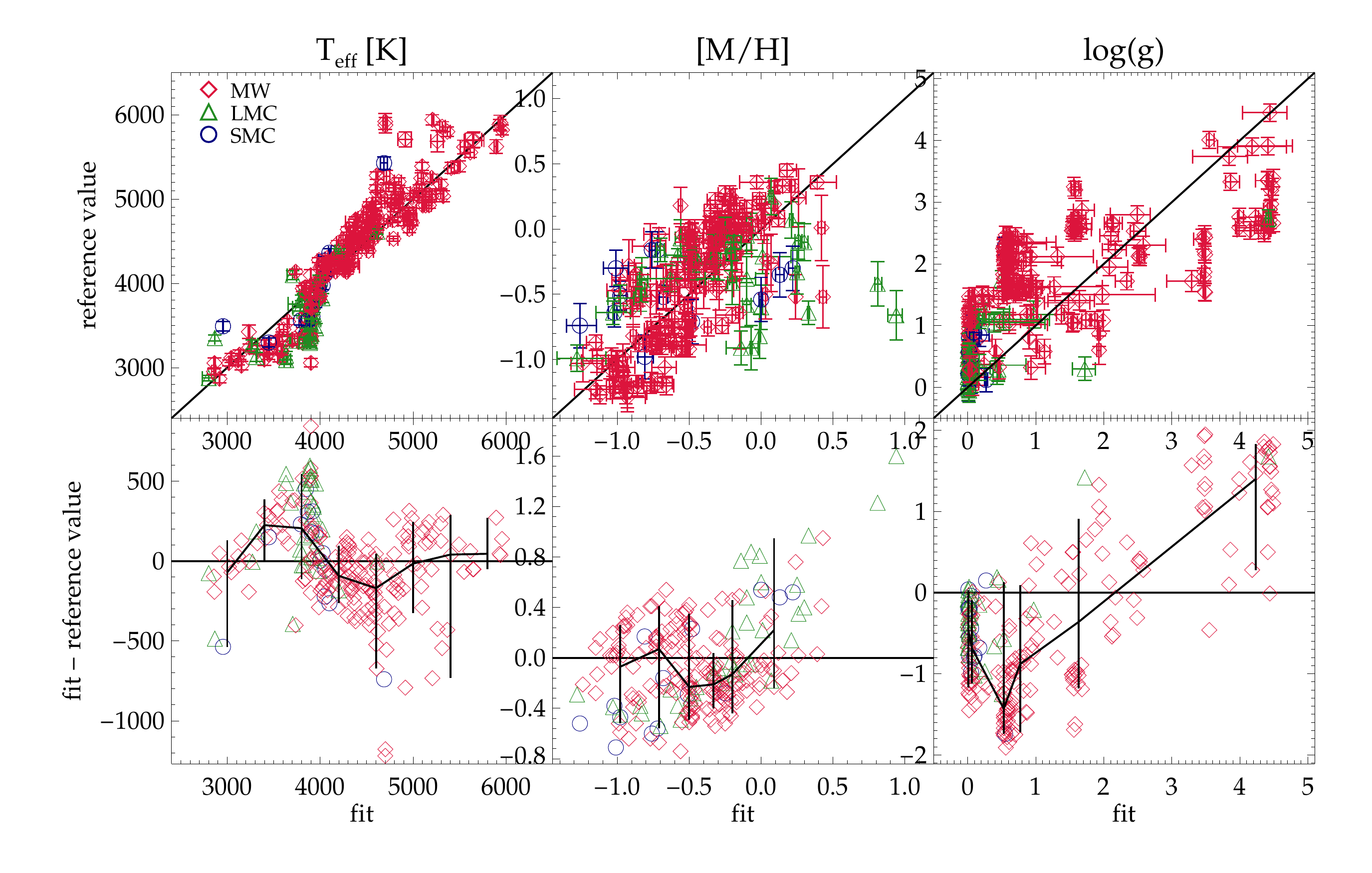}
 \caption{Comparison of the best-fitting stellar parameters obtained with \textsc{starkit} and the reference values by \citet{2019A&A...627A.138A}, for \teff\space (left), \mh\space ([Fe/H] for reference, middle), and \logg\space (right). Upper panels: fit results plotted against reference values, lower panels: fit results plotted against residual (fit results - reference values). Black coloured symbols denote the median and 0.84-percentile of the residuals. The coloured symbols denote the host galaxy: red diamonds for Milky Way (MW), green triangles for Large Magellanic Cloud (LMC), blue circles for Small Magellanic Cloud (SMC).}
 \label{fig:xsl}
\end{figure*}

We used the X-SHOOTER spectral library (XSL) DR 2 \citep{2014A&A...565A.117C,2020A&A...634A.133G}, which is a compilation of stellar spectra observed in the Milky Way, Large and Small Magellanic Clouds, at a spectral resolution $R$=7956 in the near-infrared \citep{2020A&A...634A.133G}. 
We selected only stars with stellar parameter measurements provided by \cite{2019A&A...627A.138A}, to make sure the reference parameter measurements are self-consistent. \cite{2019A&A...627A.138A} used full-spectral fitting with ULySS \citep{2009A&A...501.1269K} and the empirical MILES library \citep{2011A&A...532A..95F} to measure stellar parameters in the ultraviolet and visible ranges of the XSL spectra. 
We excluded stars with peculiar spectrum (HD 202851), Carbon stars (e.g. SHV 0534578-702532, Cl* NGC 121 T V8, HE 1428-1950), and spectra where slit-loss correction was not available. 
We convolved the XSL spectra to the mean KMOS spectral resolution ($R$=4000) and fit them with \textsc{starkit} and the PHOENIX model spectra in the same set-up as the KMOS spectra. The only difference is how we set the \logg\space bounds: 
We used \cite{2019A&A...627A.138A} \logg$_{A19}$\space measurements to set bounds, for stars with \logg$_{A19}$\textless 1.5\,dex we set 0\,dex\textless\logg\textless2\,dex, for 1.5\,dex$\leq$\logg$_{A19}$\textless 2.5\,dex we set 0.5\,dex\textless\logg\textless3.5\,dex, for 2.5\,dex$\leq$\logg$_{A19}$\textless 3.5\,dex we set 1.5\,dex\textless\logg\textless4.5\,dex, and for 
\logg$_{A19}\geq$3.5\,dex we set 2.5\,dex\textless\logg\textless5.5\,dex.
After the fits, we made quality cuts similar to those for the KMOS data, i.e. we only considered spectra with statistical uncertainties $\sigma_{\mh}$\textless 0.25\,dex, $\sigma_{T_\text{eff}}$\textless 250\,K, and $\sigma_{\logg}$\textless 1\,dex. Further, we selected only stars with similar stellar parameter ranges as our data set, meaning [Fe/H]$_{A19}$ and \mh$_\text{starkit}$ \textgreater $-1.3$\,dex, \logg\textless 4.5\,dex, 2800\,K \textless \teff \textless 6000\,K. For the remaining 237 spectra, we compared our stellar parameter results with \cite{2019A&A...627A.138A}, see also Fig.~\ref{fig:xsl}. 

We computed the mean difference and standard deviation (after 3\,$\sigma$-clipping), and obtained $\langle\Delta$\mh$\rangle$=--0.07\,dex, $\sigma_{\Delta \mh}$=0.32\,dex, $\langle\Delta$\teff$\rangle$=15\,K, $\sigma_{\Delta T_\text{eff}}$=272\,K, $\langle\Delta$\logg$\rangle$=--0.4\,dex, $\sigma_{\Delta \logg}$=1.0\,dex. The standard deviations exceeds the mean differences, which shows us that there is no strong overall bias to over- or underestimate the stellar parameters. We use the standard deviation of the residuals as systematic uncertainty measurement for our \textsc{starkit} fits.

Our \mh\space measurements are largely in agreement with the reference [Fe/H]$_{A19}$. Like us, \cite{2019A&A...627A.138A} did not consider [$\alpha$/Fe] in their fits. The scatter in our results may be in part due to non-zero elemental abundances, and that we measure the overall metallicity \mh, while \cite{2019A&A...627A.138A} measured [Fe/H]. The XSL stars have a wide range of chemical compositions, they are located in star clusters, the bulge and field of the Milky Way, in the Large and Small Magellanic Clouds, which likely broadens the \mh\space distribution at a given value of [Fe/H]. 
There are some stars with [Fe/H]$_{A19}$\textless 0\,dex, for which we obtained very discrepant values, with \mh\textgreater0.8\,dex. Most stars with residuals \mh--[Fe/H]$_{A19}$\textgreater 0.8\,dex are variable stars in the Large Magellanic Cloud. Thus, the discrepancy may be related to the chemical composition of these stars, and not affect our measurements of Galactic centre stars. 
None of the XSL DR2 spectra exceeds [Fe/H]=0.45\,dex, therefore we cannot assess whether we have any biases for stars with [M/H]\textgreater 0.5\,dex, and whether the systematic uncertainties increase, but we assume that the systematic uncertainties are the same at higher [M/H]. 

In Fig.~\ref{fig:xsl} we see a preference for \teff=3900--4000\,K in our fits, for stars with reference values $\sim$3200--4000\,K. On the other hand, at higher values of \teff$_{,A19}$, our \teff\space results can be too low. This is the case for two spectra of the same star (HD 188262), where our \teff\space estimate is $\sim$1,200\,K lower than the reference \teff.
Concerning \logg, we find that we tend to underestimate it at low values of \logg, and overestimate it at higher values. 
Due to the large systematic biases in \logg, we refrain from analysing a \teff-\logg-diagram of the \textsc{starkit} parameters, and limit our analysis to comparing [M/H] distributions is different regions of the Galactic centre.

\section{Table of stellar parameters}
\label{sec:tabsec}
\onecolumn
\begin{table}
\caption{Stellar parameters of our sample: Stellar identification number Id, equatorial coordinates R.A. and Dec., stellar parameters effective temperature $\teff$, metallicity $\mh$, surface gravity $\logg$, line-of-sight velocity $v_z$, and extinction corrected magnitude $K_{S,0}$. The full table is available as supplementary material.}
\label{tab:parameter}
\begin{tabular}{ccccccccccccc}
Id & R.A.& Dec. & $\teff$ & $\sigma_{\teff}$ & \mh&$\sigma_\text{\mh}$ & $\logg$&$\sigma_{\logg}$ & $v_z$&$\sigma_{v_z}$& $K_{S,0}$\\
&$ \left( ^{\circ} \right)$&$ \left( ^{\circ} \right)$&$ \left( \mathrm{K} \right)$&$ \left( \mathrm{K} \right)$&$ \left( \mathrm{dex} \right)$&$ \left( \mathrm{dex} \right)$&$ \left( \mathrm{dex} \right)$&$ \left( \mathrm{dex} \right)$&$\left( \mathrm{km}\,\mathrm{s}^{-1} \right)$&$\left( \mathrm{km}\,\mathrm{s}^{-1} \right)$&$\left( \mathrm{mag} \right)$&\\
\hline
$ 900013$&$ 266.50064 $&$ -28.891012 $&$ 3560 $&$ ^{+360}_{-360} $&$ 0.06 $&$ ^{+0.40}_{-0.40} $&$ 0.3 $&$ ^{+1.1}_{-1.1} $&$ -9.4 $&$ ^{+ 3.6}_{- 3.6} $&$ 8.55 $ \\
$ 900015$&$ 266.50885 $&$ -28.877468 $&$ 3960 $&$ ^{+270}_{-270} $&$ -0.14 $&$ ^{+0.32}_{-0.32} $&$ 0.1 $&$ ^{+1.0}_{-1.0} $&$ 68.8 $&$ ^{+ 1.8}_{- 1.8} $&$ 10.12 $ \\
$ 900018$&$ 266.50937 $&$ -28.878529 $&$ 3440 $&$ ^{+270}_{-270} $&$ 0.32 $&$ ^{+0.37}_{-0.37} $&$ 0.0 $&$ ^{+1.0}_{-1.0} $&$ 64.3 $&$ ^{+ 6.9}_{- 6.9} $&$ 8.77 $ \\
$ 900019$&$ 266.50790 $&$ -28.879641 $&$ 3650 $&$ ^{+280}_{-280} $&$ 0.33 $&$ ^{+0.32}_{-0.32} $&$ 0.0 $&$ ^{+1.0}_{-1.0} $&$ 49.8 $&$ ^{+ 0.4}_{- 0.4} $&$ 10.30 $ \\
$ 900020$&$ 266.50800 $&$ -28.878771 $&$ 3820 $&$ ^{+320}_{-320} $&$ -0.15 $&$ ^{+0.33}_{-0.33} $&$ 0.1 $&$ ^{+1.0}_{-1.0} $&$ 135.4 $&$ ^{+ 1.9}_{- 1.9} $&$ 10.76 $ \\
$ 900025$&$ 266.50781 $&$ -28.881580 $&$ 3830 $&$ ^{+270}_{-280} $&$ 0.22 $&$ ^{+0.33}_{-0.33} $&$ 0.5 $&$ ^{+1.1}_{-1.1} $&$ 22.0 $&$ ^{+ 1.1}_{- 1.1} $&$ 10.56 $ \\
$ 900026$&$ 266.51166 $&$ -28.877937 $&$ 3600 $&$ ^{+270}_{-270} $&$ 0.33 $&$ ^{+0.35}_{-0.35} $&$ 0.0 $&$ ^{+1.0}_{-1.0} $&$ 108.1 $&$ ^{+ 1.3}_{- 1.3} $&$ 10.44 $ \\
\\
\hline
\end{tabular}
\end{table}

\bsp

\label{lastpage}

\end{document}